\documentclass[twocolumn,epsfig,pre,showpacs]{revtex4}
\usepackage{graphicx}
\usepackage{graphics}
\usepackage{amsmath}
\usepackage{amssymb}
\usepackage{hyperref}
\usepackage{latexsym}

\def\be{\begin{equation}}
\def\ee{\end{equation}}
\def\bea{\begin{eqnarray}}
\def\eea{\end{eqnarray}}

\def\e{\epsilon}

\def\f{\frac}

\begin{document}
\title{Elastoplastic theory for  the dynamics of
solid-solid transformations : role of non-affine deformation in microstructure selection}

\author{Jayee Bhattacharya$^1$, Arya Paul$^1$, Surajit Sengupta$^1$}
\affiliation{$^1$Unit for Nano-Science and Technology, S.N. Bose National Centre for Basic Sciences, Block JD, Sector III, Salt Lake, Calcutta 700 098, India}
\author{Madan Rao$^{2,3}$}
\affiliation{$^2$Raman Research Institute, C.V. Raman Avenue, Bangalore 560 080, India\\
$^3$National Centre for Biological Sciences (TIFR), Bellary Road,
Bangalore 560 065, India
}

\begin{abstract}
We study the nucleation dynamics of a model solid state transformation and the criterion for microstructure selection using a coarse-grained molecular dynamics (MD) simulation. Our simulations 
 show a range of microstructures depending on the depth of quench. We closely follow the dynamics of the solid and find that transient {\em non-affine zones} (NAZ) are created at and evolve with the rapidly moving transformation front. The dynamics of these plastic regions determines the selection of microstructure. We formulate an {\it elastoplastic theory} which couples the elastic strain to the non-affine deformation, and recover all the qualitative features of the MD simulation. Using this theory, we construct a dynamical phase diagram for microstructure selection, in addition to making definite testable predictions.
\end{abstract}

\maketitle

\section{Introduction}
The dynamics following a quench across a solid state structural transition, rarely takes the
solid to its equilibrium state\cite{hasen}. 
Severe dynamical constraints experienced by the product 
{\it inclusion} within the parent crystal, determine the 
mode of nucleation and of subsequent growth. Often, solids
get stuck in long-lived  {\it microstructures},
which depend on the depth of quench and cooling rate\cite{stuck}. 
For example, transformations occurring at high temperatures
are typically accompanied by large-scale rearrangements of atoms; in this case
the elasticity of the solid plays only a minor role in determining 
microstructure\cite{hasen,stuck,cahn}. On the 
other hand, at low temperatures, only local rearrangements of atoms are possible; the 
resulting microstructures are largely determined by 
elasticity\cite{hasen,stuck}.
These are just two of the myriad possibilities explored by the transforming 
solid. Which of these is actually selected; in other words, can we construct a dynamical 
phase diagram?

In a set of papers\cite{drop,jpcm,ourprl}, we had 
 explored these issues using an MD simulation\cite{ums} of a
model system undergoing a two dimensional square to rhombic structural 
transformation. 
We found that when the transformation proceeds at a high temperature, the 
resulting product nucleus is isotropic and polycrystalline,  while a low 
transformation temperature induces the formation of an anisotropic nucleus,
roughly elliptical, consisting of a pair of twin-related 
crystallites\cite{ourprl}. The two modes of nucleation may be denoted {\it Ferrite} and 
{\it Martensite}, borrowing terminology from the microstructure of 
steel\cite{hasen}. By following the nucleation dynamics in `microscopic' detail, we had established that the ferrite nucleus is formed following extensive  
rearrangements of atomic coordinates, while the martensite nucleus follows from 
a transformation where the local connectivity of the lattice is, to a large
extent, preserved. This is consistent with the two paradigms commonly 
described in real materials. However,
these two limits are not mutually exclusive\cite{ourprl}; indeed 
for intermediate temperatures, the transformation proceeds  
such that both mechanisms may operate at different spatial and temporal 
locations\cite{ourprl}, a feature observed in real materials\cite{bhad}.
Further,  the different microstructures (twinned and un-twinned) 
were obtained simply by
 tuning appropriate kinetic parameters\cite{ourprl}, as observed in 
the heterogeneous nucleation of colloidal crystals\cite{paltwin}. These observations underline the
need for a unified theory of microstructure selection describing the dynamics of nucleation of both the ferrite and martensite and the conditions in which these microstructures obtain.

Our preliminary attempts at this unifying picture\cite{ourprl} were based on the recognition (from the MD simulation) of the role played by {\it non-elastic} variables, which we identified with local density
fluctuations. We showed that the coupled dynamics of density fluctuations and elastic strain determined the microstructure of the growing nucleus\cite{ourprl}.
Here, we provide a more refined theory 
of solid-state nucleation and microstructure selection.
Based on our MD simulations, we formulate an {\it elastoplastic} 
theory which interpolates between ferrite and martensite as 
some relevant parameter is tuned. We discuss the dynamical 
origin of microstructure selection and exhibit a dynamical phase
diagram\cite{ourprl} of the final microstructure.

\subsection{Average compatibility : geometrical versus strain-only theories}

Geometrically, a martensite results from the requirement that the product structure is obtainable from the 
parent lattice ({\em austenite}) at any time $t$ using a locally affine transformation, {\it viz.},
$$
{\bf R^{\prime}} = \mathbb{T}({\bf R},t){\bf R}
$$
where ${\bf R}$ and ${\bf R^{\prime}}$ are lattice vectors in the parent 
and product 
structures, respectively. The constraint that the martensite product {\em smoothly develops} 
from the parent phase, requires that they coexist on either 
side of a locally 
planar interface. This translates to the requirement that the 
transformation $\mathbb{T}$ possesses such a plane on which an arbitrary vector 
remains untransformed\cite{geom,hasen} -- the constraint of {\em rank one 
compatibility}, which is usually impossible to satisfy for any single $\mathbb{T}$.
Nevertheless, it may be possible to obtain an `approximate' planar interface between the 
parent and product structures when two or more degenerate variants (twins) of 
the transformation are used (Fig. 1). This interface possesses 
only {\em average} rank one compatibility\cite{geom} so that the rotation
of an interfacial vector, {\it averaged over a coarse graining length along the interface} vanishes. Given  a $\mathbb{T}$, crystallographic theories\cite{geom}
of martensite structure list all possible interfaces with average rank one 
compatibility (also known as an ``invariant plane strain'').
Note that at distances smaller than the size of the twins, the interface 
is {\em not} planar; at this scale the transformed region is in general 
not obtainable by an affine transformation of the parent. Mathematically, one may define a series of interfaces with rank one compatibility over ever finer scales using gradient Young
measures\cite{geom,kaubook}. The convergence of this series cannot be determined within geometrical theory alone, and needs an additional physical input. By including the interfacial energy cost for creating twin variants, the series of interfaces may be made to converge in the mean to a limiting
interface over the Young measure. 

\begin{figure}
\begin{center}
\includegraphics[width=7.0cm,angle=0]{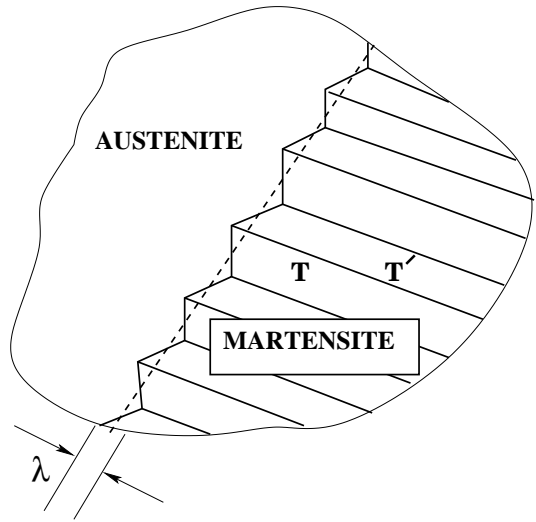}
\end{center}
\caption{
Schematic of the austenite-martensite interface : the (almost) planar interface (dashed
line) separates the austenite from the
martensite, which consists of two variants denoted by the transformations
$\mathbb{T}$ and $\mathbb{T}^{\prime}$. 
Elastic compatibility at this interface 
is restored when 
the atomic coordinates implied by this diagram are coarse-grained over 
distances of order $\lambda$, comparable to the width of the twin variants.}
\label{schematic}   
\end{figure}

A parallel approach to the study of martensitic structures was initiated 
by Barsch and Krumhansl\cite{krum}. In this framework, the martensitic structure
results from the minimization of a non-linear, elastic 
free-energy functional where the components of elastic strain are used as 
order parameter (OP) distinguishing parent and product. Unlike geometrical theories however, this programme
can be developed to study the dynamical evolution of microstructure\cite{good,strn-only,lqchen,rajiv}.
The driving force for the nucleation dynamics of martensite from the austenite is derived from the same free-energy functional, written in terms of a dynamical elastic strain tensor. 
 The free-energy functional 
supports several degenerate minima corresponding to the different variants of 
the product. Unlike the geometrical theories, these `strain-only' theories impose {\em exact elastic compatibility at all space and time}, restricting   
the elastic displacements ${\bf u}({\bf R},t) \equiv {\bf R^{\prime}(t)-R}$ to smooth, 
single-valued functions. This local constraint 
automatically disallows all configurations with defects and regions of plasticity and assumes
that the transformation is affine over {\em all} length and time scales.
Integrating out the non-order parameter (NOP) strain components, introduces 
non-local interactions between regions with different variants 
of the product, leading to complex twinned microstructures.   
  
  The exact compatibility of strain-only theories seems to be at variance with the average
compatibility of geometrical theories, and indeed with experimental evidence from
micrographs of the  austenite-martensite interface\cite{nishiyama}. One of the consequences of our
study will be to reconcile these two notions of compatibility.

To recognise this as a problem, let us recall the 
traditional hydrodynamic description of the long length and time scale dynamics of a solid.  This hydrodynamic description is written in terms of conserved variables (such as
mass density and momentum density) and broken symmetry variables (such as the displacement 
field or alternately, the strain tensor)\cite{MPP,ChaikinLubensky}. The strain-only theories\cite{strn-only}, go a level further, in declaring that the local dynamical evolution of a solid is solely via affine deformations of the parent solid. The local compatibility constraint follows naturally from this restriction on the space of allowed configurations explored by the solid, since configurations involving not only
 non-affine deformations, but vacancies, dislocations and other defects are disallowed from the dynamics.
With this restriction, the transforming solid has no choice but to be a martensite.
Strain-only theories are  therefore incapable of describing the ferrite or a host of 
other microstructures with intermediate characteristics where non-affine 
transformations are involved\cite{bhad}.

In order to provide a unified description of the dynamics of solid state transformations, we will need
to break this impasse. One way of achieving this, is to recognise that the dynamical selection of microstructure happens extremely fast, $\sim$ tens of ps, following a quench across a structural transition. 
There is therefore no fundamental reason to restrict the dynamical variables to hydrodynamical variables.
The coupling between the dynamics of these faster degrees of freedom and the conventional hydrodynamical
variables could in principle give rise to a criterion for microstructure selection. The problem is - what constitutes the relevant {\it faster degree of freedom}. In the absence of an apriori fundamental principle, we will use our coarse-grained MD simulation to explore all possible deformations left out by the earlier strain-only theories, namely,  non-affine deformations, vacancies and 
dislocations. Of these candidate fast-variables, we will see from our MD 
simulations, that the most relevant ones are transient and localised 
non-affine deformations, which inevitably accompany the tranfomation regions. 

An extension of strain-only theories to include the dynamical effects of non-affine deformations of the parent lattice, as outlined in Sect. III, results in (i) a proper theory of microstructure selection and (ii) elastic
compatibility {\it only when averged over a coarse graining distance of the order of the size of the 
twin domains}. This feature is present in our MD simulations\cite{ourprl}. This represents a reconciliation
between strain-only and geometric theories. Indeed, as we show here,
elastic compatibility, and only on the
average, {\it emerges} from a dynamical theory unifying the nucleation of
a ferrite and martensite.


\subsection{Elastoplastic description : microstructure selection and resolution of compatibility}

 Non-affine or plastic deformations
have been the subject of much study in solids subject to large shear deformations.
The many approaches to the study of plasticity include the phenomenological
 {\it elastoplastic} theories\cite{elastop,plasticity} and the more `fundamental'  non-affine field theories due to Falk and Langer\cite{langer} and Lema\^itre\cite{lemaitre}. Central to these
approaches, is the decomposition of the 
total strain into elastic and plastic parts. A plastic strain develops once the local stress exceeds a 
yield stress. 
In the elastoplastic theories of shear deformed solids, 
the yield stress and the dynamical constitutive relations between stress and
deformation rate (plasticity), are phenomenologically introduced or `derived' from models of interacting dislocations\cite{elastop,plasticity}. The more recent non-affine field theory 
approach\cite{langer,lemaitre} attempts at a unified description of plasticity in crystalline and amorphous solids, in terms of  microscopically defined shear transformation zones (STZ) representing local regions with high non-affine deformation. The coupled dynamics of STZs and elastic strain produces a `first-principles' description of plasticity, yield and work hardening which is consistent with the phenomenological elastoplastic theories.

We follow a similar program in our study of microstructure selection in solid state transformations. Our approach is guided by MD simulations of
the nucleation dynamics of the model solid introduced in \cite{jpcm,ourprl}. Our MD simulations show that {\it internal stresses} generated during the transformation, create local {\em non-affine zones} (NAZ) beyond a threshold stress. Within these zones the affine connection 
between the parent and product lattices -- and the smoothness of the displacement field 
taking the parent crystal as reference -- breaks down so that the product crystal cannot be 
described as a purely elastic distortion of the parent. One may also regard NAZs as regions of 
high dislocation density though this description is not particularly useful because at such 
high defect densities, the identity of individual dislocations is lost. 
We find that the dynamics of these NAZs determines the selection of  microstructure. We then highlight four generic principles derived from our MD  simulations, and use these to construct an elastoplastic theory for the dynamics of solid state transformations, in terms of a non-order parameter (NOP) plastic strain describing the viscoplastic nature of NAZ.  We show that this elastoplastic theory successfully describes both the ferrite and martensite nucleation and microstructure, and display a nonequilibrium phase diagram. While describing the dynamics towards a martensite microstructure, our theory  reduces to a variant of the strain-only theory\cite{strn-only}, {\it when the austenite-martensite interface is coarse-grained over a length scale $\lambda$ of the order of 
the typical size of the NAZs}. In this way, {\it average compatibility emerges from the dynamics} describing the martensite.
We follow this up with a study\cite{jayeetraj}, which associates the NAZs obtained in our MD simulations, with distinct particle trajectories. 

The rest of the paper is organized as follows. In the next section, we describe
our MD simulations on the square to rhombic transition, with special emphasis 
on the identification of non-affine zones during the solid state transformation. 
We show how local yielding 
is associated with a change in the resulting microstructure. In section III, 
we develop an elastoplastic theory of solid-state transformations and 
apply it to the specific case of the square to rhombic transition. We show 
how our coarse-grained theory, qualitatively reproduces the main features of the
MD simulations. In section IV, we discuss some implications of our study, list some
unresolved questions and indicate directions of future work.  

\section{Molecular Dynamics simulations of a model solid state transformation}
There are many real two dimensional systems which show structural 
transitions. These include confined molecular and colloidal solids\cite{sq-tr1},
flux lattices\cite{sq-tr2}, skyrmions in fractional quantum Hall 
systems\cite{sq-tr3}, magnetic colloidal particles\cite{sq-tr4} and 
colloids in electric fields\cite{yeth}. Our aim is not to 
mimic any of these systems in detail, but to construct a generic 
model which is able to describe such lattice transformations in two dimensions
using a coarse-grained potential. In addition, we would like to be able to tune the jump 
in the order parameter strain across the transition by varying the parameters in the model
potential.

A simple effective model which shows transitions between square and rhombic (a 
special case of the more general {\em oblique}) lattices,
comprises of particles which interact via
the potential\cite{ourprl,jpcm}, 
\begin{equation}
1/2\sum_{i\ne j}V_2({\bf r}_{ij})+
1/6\sum_{i \ne j \ne k} V_3({\bf r}_{i},{\bf r}_{j},{\bf r}_{k}), 
\end{equation}
where ${\bf r}_i$ is the
position vector of particle $i$, and $r_{ij} \equiv \vert {\bf r}_{ij}\vert \equiv 
\vert{\bf r}_j-{\bf r}_i \vert$. The {\it anisotropic} two-body potential\cite{jpcm},
is purely repulsive and short ranged, 
\begin {equation}
V_2({\bf r}_{ij})=v_{2}\left(\frac {\sigma_0}{r_{ij}}\right)^{12} \lbrace 1 + \alpha \cos^{2} 2\theta_{ij}\rbrace 
\label{2bdy}
\end{equation}
where $\sigma_0$ and $v_{2}$ set the units of length and energy,
$\alpha$ is an `anisotropic lock-in' parameter\cite{jpcm}, and $\theta_{ij}$
is the angle between ${\bf r}_{ij}$ and an arbitrary external axis. 
The short-ranged three-body interaction\cite{still}, 
\begin{equation}
V_3({\bf r}_{i},{\bf r}_{j},{\bf r}_{k})=v_{3}\left[f_{ij} f_{jk} \sin^{2}4\theta_{ijk} + {\rm permutations}\right]\,\, ,
\label{3bdy}
\end{equation}
where the function $f_{ij} \equiv f(r_{ij}) = (r_{ij}- r_{0})^2$ for
$r_{ij}< r_0 = 1.8 \sigma_0$ and $0$ otherwise and the angle $\theta_{ijk}$ is the 
angle between the vectors ${\bf r}_{ij}$ and ${\bf r}_{jk}$.
The two-body and three-body interactions favor rhombic and square
ground states, respectively. Inclusion of the two-body anisotropic lock-in
parameter $\alpha$ is a device to vary the jump in the order parameter from 
strongly first order ($\alpha = 0$) to a continuous transition for 
$v_3 = 0, \alpha \sim 1.5$. 
The unit of time is $\sigma_0\sqrt{m/v_2}$, where $m$
is the particle mass. Using typical values, this translates to an   
MD time unit of $1 ps$. Knowing the individual particle MD trajectories allows us to  
project time dependent atomic positions into time varying coarse-grained 
{\em fields} whose evolution can be monitored during the transformation.

Both the two and three -body potentials are purely repulsive and therefore the system needs to 
be confined either in a box of fixed volume or by an external 
compression\cite{ums}. In this paper, we discuss our results for MD simulations
in the constant number, volume (and shape), and temperature (NVT) ensemble with
periodic boundary conditions using a Nos\'e-Hoover thermostat. We have, in 
addition, carried out extensive simulations in the constant stress (N$\Sigma$T)
ensemble  with open boundaries using an additional confining 
potential which, at the same time, allows for changes of overall shape of the 
crystal during the transformation. Our main results concerning the dynamics and 
mechanism of microstructure selection are the same in both the ensembles. A 
detailed comparison of MD simulation of  our model system in various 
ensembles, starting from a variety of initial states and for the full range
of the potential parameters is being prepared for publication elsewhere.    

\begin{figure} 
\begin{center}
\includegraphics[width=7.0cm,angle=0]{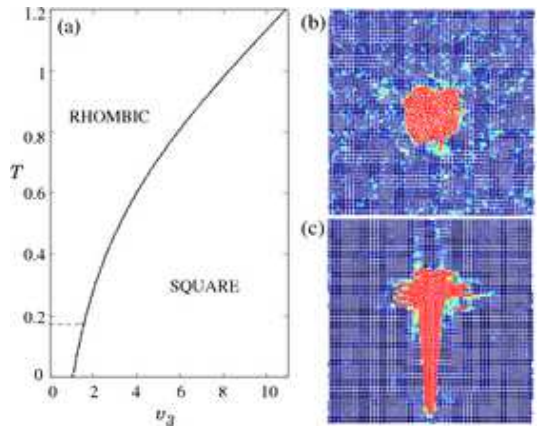}
\end{center}
\caption
{(color-online) (a) Phase diagram in the $T-v_3$ plane (with $\alpha=0$), the 
solid line is the phase boundary between square and rhombic crystals. The dashed line marks the temperature above which anisotropic,
twinned nuclei become rare and is identified as the martensite-start
(or $M_s$) temperature\cite{hasen} for our model. Typical product 
nuclei formed following quenches at (b) $T=0.8$ ($v_3 = 10 \to 5.5$), to obtain the isotropic `ferrite' and (c) $T=0.1$ ($v_3 = 5 \to 1.65$) to obtain the anisotropic, twinned martensite, starting with an 
equilibrated square parent crystal composed of $12099$ particles.}
\label{phdia}
\end{figure}

The equations of motion for up to $N=20000$ particles are integrated using a 
Verlet scheme\cite{ums} with a time step $\Delta t = 10^{-3}$\cite{dynam}. An 
accurate equilibrium phase diagram of the system (Fig.\ref{phdia})
in the T-$v_3$ plane for density $\rho_0 = N/V = 1.1$ is obtained 
by computing and comparing the free energies of square and rhombic lattices
using the technique outlined in \cite{ho}. 

We have now set the stage for a detailed study of the nucleation dynamics of a solid in solid,
following a quench across the structural transition. Our effort will be to extract general matters of principle from these simulations; we will highlight these as we go along. These principles will form the basis for the elastoplastic theory of solid state transformations (section III).

A typical quench from a square to a rhombic solid into a region where the square lattice
is metastable, initiates multiple nucleation events (at least at high temperatures), making a quantitative analysis of
the dynamics of a single critical nucleus, cumbersome. We get over this difficulty by introducing a
nucleation seed at the center of the simulation box.
The seeding consists of replacing the
central particle with a particle whose size $\sigma$ is smaller by a factor $\delta \equiv (\sigma_0-\sigma)/\sigma_0$, we have taken $0.25 \leq \delta \leq 1$, so as to obtain nucleation events within reasonable computation time. 
Having equilibrated the seeded square crystal at large $v_3$, we ``quench'' 
across the phase coexistence line by varying the coefficient of the three-body term $v_3$ at two different temperatures $T=0.8$ and $T=0.1$\cite{note-quench}. While the seeding
is a matter of convenience at the higher temperature, it is necessary at the lower temperature.
The transformation at the lower temperature proceeds via heterogeneous nucleation\cite{ourprl}.

In Fig.\ref{phdia}(b),(c), we show a snapshot of the resulting microstructure following a quench 
from the equilibrated square lattice at temperatures $T=0.8$ and $0.1$, 
respectively. The colors indicate the local bond-angle order parameter which 
is defined to vary from $0$ (blue) in the square lattice to $1$ (red) in 
the rhombic\cite{ourprl}. 
It is clear from the particle 
position snapshots, Fig.\ref{phdia}(b),(c), that the 
product nucleus is isotropic for large temperatures and highly anisotropic 
for small temperatures. We identify the isotropic nucleus with a {\em ferrite}
and the anisotropic one with {\em martensite}\cite{ourprl}. This identification is reinforced by
showing that the latter is twinned.\\

1. {\it Solid state transformations predominantly proceed via nucleation. At low temperatures, the nucleation of the product solid is heterogeneous and is initiated by `seeding' the parent}.\\

To follow the dynamics in quantitative detail, we compute the coarse-grained local strain field using
the procedure introduced in \cite{langer}. 
Briefly, we compare the immediate neighborhood $\Omega$, centered around ${\bf r}$, of any tagged 
particle $0$ (defined using a cutoff distance equal to the range of the potential) 
in the initial, 
reference, lattice (at time $t=0$) with that of the same particle in the 
transformed lattice. We obtain the ``best fit'' local affine strain 
$\e_{ij} = \mathbb{T}_{ij} - \delta_{ij}$ ($\delta_{ij}$ is the
Kronecker tensor) which maps as nearly as possible all the particles $n$ in 
$\Omega$ from the reference to the transformed lattice using an affine 
connection. This is
done by minimizing the (positive) scalar quantity,
\begin{eqnarray}
D^2_{\Omega}({\bf r},t) & = & \sum_{n \in \Omega} \sum_i \lbrace r_n^i(t) - r_0^i(t) - \sum_j (\delta_{ij} + \e_{ij}) \nonumber \\
        &  &  \times (r^j_n(0) - r^j_0(0)) \rbrace^2 
\label{nonaffine}
\end{eqnarray}
with respect to choices of affine $\e_{ij}$.  Here the indices 
$i\,{\rm and}\,j = 1,2$ (or $x,y$) and $r^i_n(t)$ and $r^i_n(0)$ are the 
$i^{th}$ component of the position vector of the $n^{th}$ particle in the 
reference and transformed lattice, respectively. Any {\em residual} value of 
$D^2_{\Omega}({\bf r},t)$ is a measure of {\em non-affineness}.
\begin{figure}
\begin{center}
\includegraphics[width=7.0cm,angle=0]{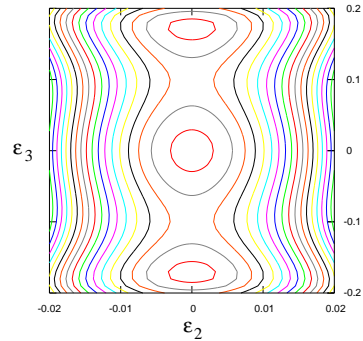}
\end{center}
\caption
{(color-online) Contour plot of the zero temperature energy per particle for
$\rho = N/V = 1.05, \alpha =  1$ and $v_3 = .204$ as a function of the
OP strains $(e_2,e_3)$ showing a metastable square minimum at $(0,0)$ and two degenerate, 
stable rhombic ({\em oblique}) minima at $(0,\pm.18)$.}
\label{tzero}
\end{figure}

The rhombic lattice is a special case of the general oblique lattice -- one of the 
five possible two dimensional Bravais lattices. In general, we need two
order parameters (OP) to describe the transition between 
square and oblique lattices, i.e., between the space groups $p4mm \to p2$.
These are the affine shear strain $e_3 = \e_{xy}=\e_{yx}$ 
and the deviatoric strain $e_2 = (\e_{xx}-\e_{yy})$. Thus from 
symmetry considerations alone, we would expect to obtain four symmetry related product
phases\cite{Avad-sym}.
However, the microscopic model used by us obtains a rhombic lattice for which $e_2$ 
identically vanishes and the four equivalent variants merge in pairs to give {\em two} 
symmetry related products. To show this we have plotted in Fig.\ref{tzero} the 
$T=0$ energy obtained for our model solid with the parameters $\alpha = 1, v_3 = 0.2$ and
at $\rho = 1.05$ for various values of $e_2$ and $e_3$ taking the square lattice as the 
reference. Apart from the minimum corresponding to the square lattice 
we obtain only two other minima representing the two rhombic variants. The value of $e_2$ 
at all the three minima is zero. It is therefore sufficient to use $e_3$ as the sole 
order parameter (OP) for this transition\cite{note-OPjump}.

\begin{figure}
\begin{center}
\includegraphics[width=6.5cm]{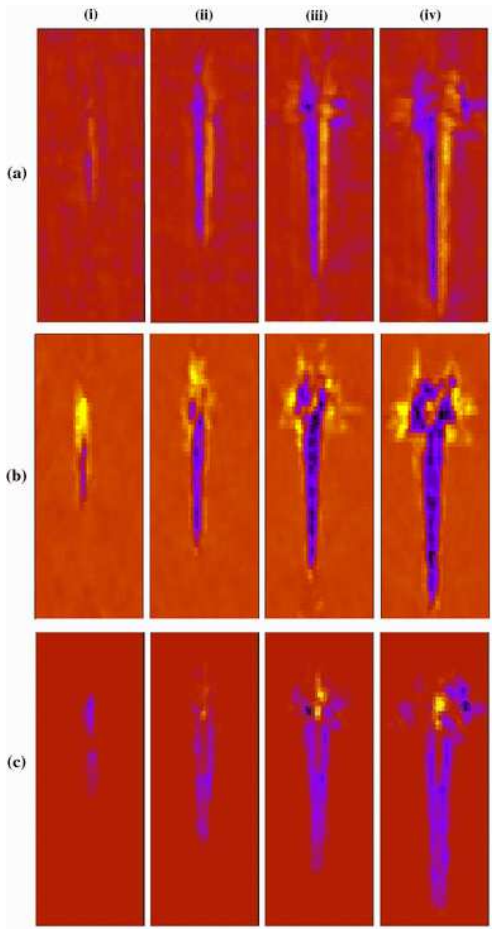}
\end{center}
\vskip -.5 cm
\caption
{(color-online) Best-fit affine strains and residual non-affine deformations,
obtained from MD simulations of particles 
undergoing a square to rhombic transition at $T=0.1$, for time slices (i) $2000$, (ii) $3000$, (iii) $4000$ and (iv) 
$5000\,\Delta t$. Plots obtained by coarse-graining the $N=110\times110$ lattice to a $64\times64$ lattice. (a) Order parameter (shear) strain $e_3$, colors 
show $e_3$ from $-0.3$ (black) to $0$ (brown) to $0.3$ (yellow). 
The twinned microstructure is  clearly visible, even at 
earlier times. (b) Non-order parameter (volumetric) strain $e_1$, colors show 
$e_1$ from $-0.5$ (black) to $0.3$ (yellow). The equilibrium value of $e_1$ 
is nonzero within the rhombic phase, in addition, $e_1$ appears at the two 
ends of the twinned microstructure due to elastic coupling to the order 
parameter $e_3$. (c) Non-affine deformation $\chi$. Colors show  $\chi$ 
ranging from $-1$ to $1$. Note that $\chi \to 0$ at the centre of the 
growing nucleus at large times. The NAZs surround the 
growing nucleus and are created at and advected by the front.  Jammed and 
unjammed NAZs occur at the `top' and `bottom' of the nucleus, respectively, 
sharing the same spatial symmetries as $e_1$.
} 
\label{martsim}
\end{figure}

Figure \ref{martsim}(a) shows the nucleation and growth of the twinned martensite nucleus, following the lower temperature quench -- we have 
plotted the best-fit $e_3$ for snapshot configurations of $N=110\times110$ 
particles 
at time steps of $2000$, $3000$, $4000$ and $5000$ $\Delta t$. 
The twinned structure of the nucleus composed of the two degenerate rhombi 
(characterized by positive and negative values of $e_3$, separated by a sharp boundary) is evident even at the earliest time, and becomes
more pronounced as time progresses. The constraint of fixed density forces a dynamical coupling between the affine OP 
strain and the affine non-order parameter (NOP) volume strain 
$e_1 = \e_{xx}+\e_{yy}$, so that the transformation is also accompanied by a 
volume change, Fig.\ref{martsim}(b). 
 As a result, as the transformation proceeds, more and more 
particles are pushed up against the surrounding untransformed square
lattice which creates a {\em jammed} region at one end and an {\em unjammed}
region at the other end of the anisotropic martensitic nucleus, Fig.\ref{martsim}(b). 
\\

2.  {\it The dynamics of transformation is described by an affine OP strain (here, shear strain) characterizing the microstructure of the growing nucleus, and an affine NOP strain (here, volumetric strain), which is slaved to the former}.\\

We can now use the residual $D_{\Omega}^2$, (\ref{nonaffine}) to extract the spatio-temporal variation of any non-affine deformation that is produced during the
transformation. To be able to distinguish between non-affineness arising from different components of the strain (shear or volumetric) distortion, we need to incorporate the notion of jamming in the definition of non-affineness (\ref{nonaffine}). In the context of granular compaction\cite{edwards} and glassy materials\cite{liunagel}, jamming has been quantified in terms of changes
 in the local free-volume relative to the reference state. In our context, this translates into computing the
 relative change in the distance
between particles within $\Omega$ in the direction of motion of the particles 
in the nucleus, denoted by $\Delta l$; we may thus define a quantity
$\chi({\bf r},t) = -D^2\,{\rm sign}(\Delta l)$, which takes both positive (jammed) and negative (unjammed)
values. For the martensite nucleus, the jammed and unjammed non-affine zones (NAZs) are shown in Fig.\ref{martsim}(c); as the transformed region grows, $\chi$ is localized and 
advected by the transformation front. Note that the spatial symmetries of $\chi$ are the same as that of $e_1$ at all times (compare Figs.\ref{martsim}(b) and (c)), and so we associate the non-affineness predominantly with the NOP or volumetric strain. This restriction of non-affineness to the NOP strain alone, could be specific to the square-to-rhombus transition; in transitions between other structures, there could be a fair degree of plasticity associated with the OP strain too. We will return to this point in section IV.
Consistent with geometrical theories, the NAZs are absent at the twin interface; this interface is coherent and the twins are simply related to each other by an affine transformation.

\begin{figure}
\begin{center}
\includegraphics[width=7.0cm,angle=0]{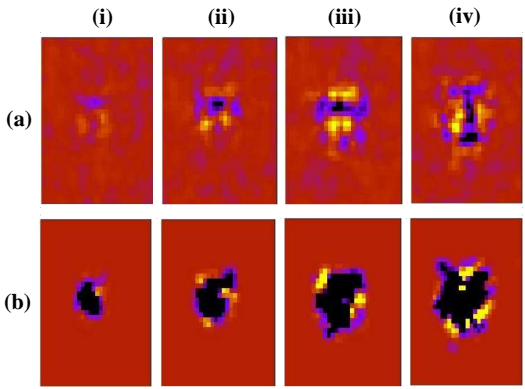}
\end{center}
\caption
{(color-online) Best-fit affine strains and residual non-affine deformations,
obtained from MD simulations for a quench at $T=0.8$, for time slices (i) $8000$, (ii) $10000$, (iii) $13000$ and (iv) 
$15000\,\Delta t$. Same coarse-graining as in Fig.\ref{martsim}.
(a) Order parameter (shear) strain $e_3$, colors show $e_3$ from $-0.3$ (black) to 
$0.3$ (yellow). Note that unlike the martensite, no clear spatial 
pattern in $e_3$ can be discerned. The local structure of the product 
nucleus is polycrystalline, with individual grains which coarsen 
with time. 
(b) Non-affine deformation $\chi$,
colors show  $\chi$ ranging from $-1$ to $1$. Note that non-affine 
regions are present throughout the interior of the nucleus, signifying extensive
plastic deformation during growth.}
\label{ferritesim}
\end{figure}

In contrast, Fig.\ref{ferritesim}(a) shows the nucleation and growth of the ferrite nucleus, following the higher temperature quench -- as above, we have 
plotted the best-fit $e_3$ for snapshot configurations of $N=110\times110$ 
particles at time steps of $8000$, $10000$, $13000$ and $15000$ $\Delta t$. 
The nucleus is composed of polycrystalline grains of the rhombic phase separated by large angle grain boundaries. As time progresses, the grains rotate with respect to each other, giving rise to large
non-affine distortions {\it even in the bulk of the nucleus}. This is reflected in the large values of $\chi$ in the bulk of the growing nucleus, Fig.\ref{ferritesim}(b). However, a spatial average of the instantaneous values $\chi$ and $e_3$ over a scale larger than the grain size, gives zero for both. Similarly, a time average of the local $\chi$ and $e_3$ over a window corresponding to typical grain reorganization times, gives zero for both.  The plastic zone spreads throughout the product region causing extensive 
atomic rearrangements. 

\begin{figure}
\begin{center}
\includegraphics[width=7.0cm,angle=0]{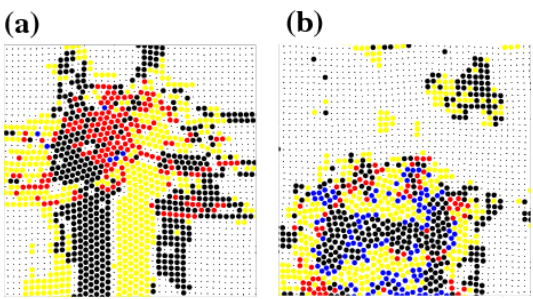}
\end{center}
\caption
{(color-online)
Close up of a region from the nucleus (a) in martensite : corresponding to 
Fig.\ref{martsim}(iii) and (b) ferrite : corresponding to  Fig.\ref{ferritesim}(iv). The color code is as follows:  
black dots -- untransformed regions; yellow and 
black circles  -- affine regions with $+$ve and $-$ve $e_3$;
red and blue circles -- jammed and un-jammed non-affine regions. 
In (a) a similar non-affine region arises 
at the other (bottom) end of the twinned region (not shown).}
\label{nonaf}
\end{figure}

We now take a closeup look at the NAZs -- Fig.\ref{nonaf}(a),(b) shows snapshots of the atomic positions in the NAZs of the martensite and ferrite nucleus, respectively. It is clear that the atomic configurations
in the NAZs are highly amorphous, with no clear relation to the reference parent lattice.
It seems meaningless to describe NAZs  in terms of a density of dislocations, since the 
reference state has no unique physical significance for characterizing the current state
in the NAZs. Even if we were to describe the state of NAZs in terms of dislocations, the 
density of dislocations would be so high as to have overlapping cores, thus rendering this language
inadequate. It is more reasonable to describe the NAZs in terms of fluctuations
in the local density $\phi({\bf r},t) = (\rho({\bf r},t)-\rho_0)/\rho_0$, where $\rho({\bf r},t) = \sum_{n \in \Omega}  \delta({\bf r}-{\bf r}_n(t))$, and $\rho_0$ is the average uniform density.
Indeed in \cite{ourprl}, we had studied the dynamics of  
$\phi({\bf r},t)$ in great detail and demonstrated its involvement with dynamics of transformation and microstructure selection. Here we find by explicit computation that $\phi$ and $\chi$ are related -- localized regions with large $\phi$ correspond to large $\chi$ and so on. \\

3.  {\it Right from its initiation, the transformation is accompanied by non-affine deformations 
primarily associated with NOP (here volumetric) strain. The dynamics of non-affine deformations determines the microstructure.}\\

We will now show that NAZs are produced when the local volumetric stress exceeds a threshold value. We compute the instantaneous local stress from our MD simulations by spatially 
averaging the generalized virial,
$$
\sigma_{ij} = \langle \sum_{n \in \Omega} {\bf F}_i {\bf r}_n^j \rangle 
$$
over cells $\Omega_M$ containing $M$ particles where $1 \ll M < N$. The choice of $M$ is 
dictated by the mutually competing considerations of proper averaging and 
obtaining information over a fine enough length scale. We have chosen 
$M=100$ as a compromise between these considerations. Further, in order to 
obtain good statistics for any time $t$, we average over many independent 
quench runs. Thus the spatio-temporal resolution of the computed ${\bf{\sigma}}$ is not as high as the one for the coarse-grained strain ${\bf e}$. Details pertaining to the evaluation of this quantity for our 
potential including three-body terms is given in Appendix A. 

\begin{figure}
\begin{center}
\includegraphics[width=7.5cm,angle=0]{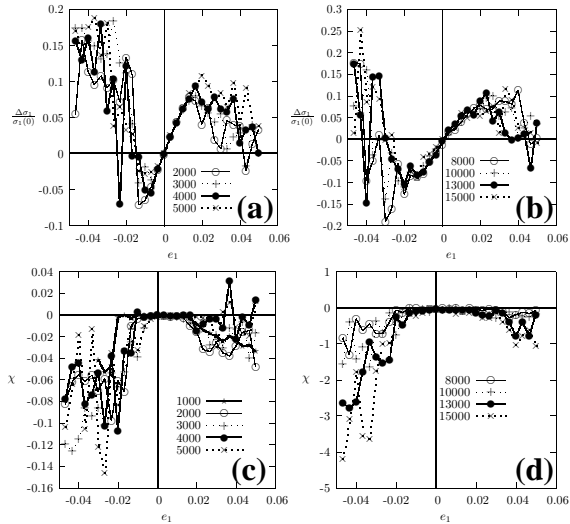}
\end{center}
\caption{Local stress expressed as a fractional difference from the volumetric
stress at $e_1 = 0$ ($\Delta \sigma_1/\sigma_1(0)$),
plotted against the local strain at different times (symbols) obtained for the 
(a) quench at $T=0.1$, averaged over $40$ independent quenches, and 
(b) quench at $T=0.8$, averaged over $35$ independent quenches. 
(c) and (d) corresponding plots of $\chi$ vs $e_1$. 
The linear Hooke-an regime  represents 
the local elastic response at small local stress. Beyond  a threshold, the system {\it yields} locally, giving rise to a nonlinear stress-strain behavior {\it and} simultaneously non-affine deformations $\chi>0$ (red symbols). The regions in real space associated with this local plastic regime are identical to the NAZs.}
\label{maryld}
\end{figure}

\begin{figure}
\begin{center}
\includegraphics[width=6.0cm,angle=0]{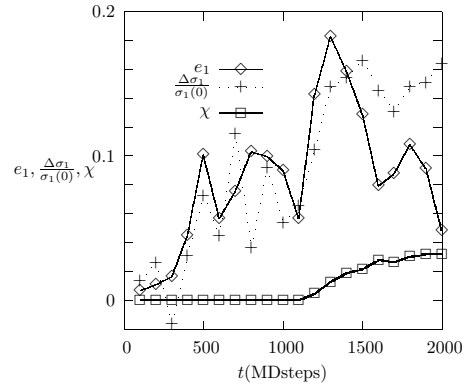}
\end{center}
\caption{Time dependence of local stress $\Delta \sigma_1/\sigma_1(0)$, strain $e_1$ and $\chi$ near the growing nucleus after a quench at 
$T=0.1$. Note that initially $\chi = 0$ and the stress (apart from large
statistical fluctuations) is proportional to strain. When the 
local stress exceeds a threshold, $\chi$ begins to increase, and the stress versus strain is highly nonlinear.}
\label{timeld}
\end{figure}

We can now compute the local volumetric stress $\sigma_1$, affine volumetric strain $e_1$ and non-affine $\chi$, averaged over the coarse-grained cell $\Omega_M$, 
at different times following the quench. This is plotted in Fig.\ref{maryld}(a),(b),
where we have expressed the local stress as a fractional 
difference about the value of $\sigma_1$ for 
$e_1 = 0$, viz., the undistorted region. The $\sigma_1$-$e_1$ plot shows a linear
elastic regime for  those coarse-grained cells where the strain $e_1$ is small; concomitantly the
non-affine $\chi$ is zero (Fig.\ref{maryld}(c),(d)). Coarse-grained cells where $e_1$ is larger than a threshold, show yielding (nonlinear
and erratic $\sigma_1$-$e_1$) and appreciable plastic flow, $\chi \neq 0$. We have verified that these coarse-grained cells showing
plastic deformation are indeed the NAZs reported above. We now focus on one 
coarse-grained cell, and study the time development of $\sigma_1$, $e_1$ and $\chi$ as the
transformation proceeds (Fig.\ref{timeld}). We find that at earlier times, the strains are
small and the stress-strain response is elastic. Beyond a yield stress $\sigma_{1c}$, the stress-strain
relation is nonlinear, giving rise to non-affine deformations $\chi \neq 0$. Following yielding, the local stress eventually decreases, often exhibiting oscillatory behavior. We find that the threshold  
stresses $\sigma_{1c}$, when expressed as a 
fraction of the ambient stress is only weakly dependent on 
temperature. \\


4. {\it Non-affine deformations are produced when the local stress crosses a 
threshold. The threshold stress is only weakly dependent on temperature.}\\

In the next section, we will use the four principles highlighted above to 
construct an elastoplastic theory for the dynamics of solid state 
transformations. These four principles are generic, and should not depend 
on the choice of potential, or the nature of the transformation.
We will show that the development of the 
microstructure is crucially influenced by the dynamics of the NOP plastic 
strain associated with NAZs.  In what follows we will
develop this theory first in a general setting and then specialize 
 to the particular case of the square to rhombic transformation studied in this section.

\section{Elastoplastic theory of nucleation dynamics of solids}

To describe the solid state structural transition with elastic and plastic strains, we first write the total strain tensor as  $\e \equiv \{ \e^A_T, \e_V\}$, where the
affine OP or transformation strain $\e^A_T$ connects the parent and product 
lattices, and the NOP strain $\e_V$ is split into an affine part and a non-affine or plastic strain, 
$\e_V = \e^A_V - \e^P_V$ ($\e_V$ can have many components, we have dropped the tensor indices for clarity). Note that our association of plasticity with the NOP strain {\it alone}, follows from
the previous section, however as we remarked, this could be special to the square-to-rhombus transformation. In general, there could be plastic deformations associated with the OP strain too; we
will comment on this in section IV.

The transformation is described by  a free-energy functional
\begin{equation}
{\cal F} = \int_{\bf r} F_T(\{\e^A_T\}) + \alpha (\nabla \e^A_T)^2 + \beta (\nabla \e^A_V)^2 
+ \frac{C}{2} \vert \Lambda \e^A_V - \e^A_T - \e^P_V \vert^2 \, ,
\label{free}
\end{equation}
with $F_T(\{\e^A_T\})$ having three minima
corresponding to the parent phase $\e^A_T = 0$ and (symmetry related) variants of the
product phase, $\e^A_T \neq 0$. This choice of ${\cal F}$ is dictated by simplicity, more
complicated forms having symmetry allowed nonlinear cross-couplings can be envisaged, but these will not change our framework. In the absence of plastic deformations, a variation of (\ref{free}) with respect
to $\e^A_V$, gives the desired relation connecting the OP and the NOP strain. The coefficients
$C$ and $\Lambda$ determine the coupling between $\e_V$ and $\e^A_T$ and the contribution
of the plastic strain to the free energy. 

The driving force for $\e^A_T$ is the chemical potential gradient 
$\delta {\cal F}/ \delta \e^A_T$ to form the rhombic phase and the dynamical equations take
the general form\cite{strn-only},
\begin{equation}
\Psi({\e_T},{\dot\e_T},{\ddot \e_T}; \e_V)  = 0 
\label{elastic}
\end{equation}

The dynamics of the affine NOP strain is slaved to the OP strain. 
This takes the form of a local force balance\cite{landau},
\begin{equation}
\nabla\cdot\sigma_V = 0 
\label{me}
\end{equation}
where the local NOP stress $\sigma_V$ is related to the
instantaneous equilibrium value of the NOP strain,
\begin{equation}
\sigma_V = \frac{\partial {\cal F}}{\partial \e^A_V} \, .
\end{equation}

The dynamical equations for the plastic NOP strain $\e^P_V$ are constructed
phenomenologically following the principles listed above. We include the
physics of threshold stress and yield flow, by a constitutive relation between 
stress and strain rate,
\begin{eqnarray}
\dot\e^P_V & = & \frac{1}{h}\,\,\,(\sigma_V - \sigma_{0})^{\frac{1}{\theta}} \,\,\,\,{\rm if}\,\,\, g(\sigma_V;\sigma_{Vc}) > 0 \nonumber \\
               & = & 0\,\,\,\,\,\, {\rm otherwise}
\label{plastrain}
\end{eqnarray}
where $h, \theta$, $\sigma_0$ and $\sigma_{Vc}$ are material parameters and 
$g(\sigma_V;\sigma_{Vc})$
is the appropriate (material dependent) threshold or yield criterion\cite{elastop,plasticity}, which in principle can incorporate history dependence. In the next section, we will make the simplified `Newtonian'
ansatz, $\theta=1, \sigma_0=0$, for this case the parameter $h$ is related to the relaxation time
of plastic flow. The ratio of the stress to strain rate, $\sigma/{\dot\e^P_V} = \eta$, is a bulk viscosity. At the yield stress, this viscosity diverges, signifying jamming. Note $\sigma$ is an internal stress, and can therefore locally decrease and increase once the solid yields, giving rise to oscillatory behavior, as seen in Fig.\ref{timeld}. The form of the constitutive relation 
\ref{plastrain} precludes the possibility of ``creep'' since $\dot\e^P_V$ 
vanishes at zero stress. This motivated by our MD results as shown in 
Fig.\ref{timeld}.  

In addition, owing to plasticity, the local affine strains do not satisfy the usual 
St. Venant's compatibility\cite{stuck} --- instead, the amount of incompatibility is 
{\it exactly accounted for by the amount of plasticity generated}. This implies that the 
local St. Venant's condition should be rewritten as\cite{zipp},
\begin{equation}
\nabla \times (\nabla \times \e)^{\dagger} = 0
\label{stven}
\end{equation}  
where $\e$ is the {\it total} strain, which includes $\e^P_V$. In regions where the local
plastic deformation is zero, this reduces to the usual St. Venant's compatibility
condition.

The equations of constraint, (\ref{me}) and (\ref{stven}), are used to express $\e^A_V$ in terms of
$\e^A_T$; equations, (\ref{elastic}) and (\ref{plastrain}), are then used with appropriate initial conditions to describe the elastoplastic theory for the dynamics of solid state transformations.

As we  highlighted in the previous section, 
the nucleation process is heterogeneous. This implies that since equations (\ref{elastic}) and (\ref{plastrain}) are deterministic, one needs to introduce `seeds' in order to initiate nucleation events. In our elastoplastic 
theory, this can be introduced as initial conditions either in the 
local strain $\e^A_T$ or the local stress field $\sigma_V$. A random (initial) distribution of the stress 
field $\sigma_V$, can produce local plastic strain $e^P_V$, if the local $\sigma_V$ seed is larger than
the yield stress. This will in turn nucleate the transformed solid, via $\e^A_T$. The results obtained from solving the dynamical
equations must be then averaged over realizations of the quenched random stress field.
A simpler strategy is to directly introduce a seed in the transformation strain, $\e^A_T$, by creating a small twinned region with
a single twin boundary (since the dynamics of $\epsilon^A_T$ is conserved). 
This allows us to follow 
the subsequent dynamics in precise detail and does not require any averaging over noise realizations.   

We use the elastoplastic model to address two separate but related issues.
We will first determine the late time {\it morphology} and {\it microstructure} of the growing
nucleus following a quench across the structural phase boundary and construct a dynamical  phase diagram akin to Fig.\ref{phdia} of our MD simulation, or Fig. 4 of \cite{ourprl}.
We compute the shape of the growing nucleus from the shape asphericity
$A = (\lambda_1 - \lambda_2)/(\lambda_1 + \lambda_2)$, where 
$\lambda_i$ are the eigenvalues of the moment of inertia tensor of the 
nucleus\cite{jayeetraj}. By quenching into different regions of the dynamical phase
diagram, we will study the dynamics by which specific microstructure gets selected
within our elastoplastic theory. 

A brief comment, before we discuss our explicit computation : strain-only theories\cite{strn-only} set $\e^P_V=0$, and use (\ref{me}) and (\ref{stven}) to eliminate $\e^A_V$ in terms of $\e^A_T$, leading to long-ranged interactions
and dissipation in $\e^A_T$, both of which are spatially anisotropic. This is ultimately responsible for producing the twinned 
microstructure of martensites in the strain-only description; the size of the twins is set by
elastic parameters alone\cite{strn-only}. These theories cannot describe the occurrence of the ferrite. In our elastoplastic description, inclusion of local plastic deformation in the form of $\e^P_V$ has two effects ---  it {\it screens} and {\it isotropises} the non-local interaction and dissipation
kernel.
This is ultimately responsible for the destruction of the twin pattern, resulting in a ferrite. Further, the size of the twins depends on elastic, as well
as plastic parameters. We provide a detailed analysis of these effects in a forthcoming publication.

\subsection{The square to rhombic transition}
We now present an explicit calculation for the particular case of the square to rhombus
transformation in two dimensions, studied in section II. As mentioned there, the square to rhombus transformation 
is a special case of the square ($p4mm$) to oblique ($p2$), and is thus described by two
order parameter strains characterising the $4$-degenerate product phases. In general, one can 
construct a Landau theory for this transition\cite{Avad-sym,toledo} using terms upto sixth 
order in the OP strains $e^A_T = \{e_2,e_3\}$ and quadratic in the NOP strain 
$\e_V \equiv e_1$. We then decompose the NOP strain into a slaved, 
affine NOP strain $\e^A_V \equiv e^A_1$, and a dynamical non-affine NOP strain 
$\e^P_V \equiv e^P_1$, enabling the total NOP strain to be written as, $e_1 = \e^A_1 - \e^P_1$. With this decomposition, we can then proceed with the
general treatment outlined in the last subsection.

However to compare with the results and phenomenology of the MD simulation of the model solid described by the microscopic potential (\ref{2bdy}),(\ref{3bdy}), it is more convenient to look at a restriction of this problem. Recall  (see, Fig.\ref{tzero}) that our microscopic potential supports {\it two} rather 
than the possible four product minima. This implies that for this choice of potential, there is therefore only 
{\it one} minimum in the 
$e_2$ direction; it thus suffices to retain upto quadratic terms in $e_2$ in the strain free-energy functional. The minimal free-energy functional, sufficient to describe this square to rhombus transition
is given by, 
\begin{eqnarray}
{\cal F}  &=&  \frac{1}{2}\int\,{\rm dxdy}\,\left[ a_1(e_1+e_1^P)^2\,+\,a_2e_2^2\,+\,a_3e_3^2 \right.    \\\nonumber
          & & \left.+\,c_1(\nabla (e_1+e_1^P))^2\, +\,c_2(\nabla e_2)^2\,+\,(\nabla e_3)^2 - e_3^4 + e_3^6 \right]
\label{frengy}
\end{eqnarray}
in terms of the OP strains $e_3$ and $e_2$ and the NOP strain $e_1$. The only other term to 
quadratic order in $e_2$, viz., $e_2^2 e_3^4$ has also been dropped since it does not
influence the phase transition. Note that this form of the free-energy functional can be 
recast just as in (\ref{free}). The three elastic constants
$a_1, a_2, a_3$ define the linear elasticity of the square phase. 
The coefficients of the quartic and sixth order terms as well as that of 
$\nabla e_3$ can be scaled to unity by rescaling $e_1$, ${\cal F}$ and the 
spatial coordinates ($x,y$). 
The coefficient $a_3$ represents the degree of under-cooling; we work in a parameter range 
where the square crystal is metastable and the rhombic crystal is stable at equilibrium.

The affine NOP strain is slaved to the OP strains; we make use of the two conditions
(\ref{me}) and (\ref{stven}), to express $e^A_1$ in terms of $e_2$ and $e_3$. Mechanical 
equilibrium implies
$\nabla \cdot {\bf{\sigma}} = 0$. The modified St. Venant condition
(\ref{stven}) reads,
\begin{equation}  
{\bf \nabla}^2 e_1 - (\nabla_x^2 - \nabla_y^2)e_2 - 4\nabla_x \nabla_y e_3 = 0 \, ,
\label{venant}
\end{equation} 
which includes both affine and plastic deformations.
The relation between the total NOP strain $e_1$ and the OP strains is most conveniently expressed in ${\bf k}-$space,
\begin{equation} 
\tilde{e}_{i}({\bf k}) = \tilde{{\cal Q}}_{i3}({\bf k})\,\tilde{e}_3({\bf k})
\label{kernel}
\end{equation}
with the kernels, 
\begin{subequations}
\begin{eqnarray}
\tilde{{\cal Q}}_{13}({\bf k}) & = & \frac{4 a_2 - 2 a_3}{a_1+a_2}\,\frac{k_x k_y}{k^2} \nonumber \\
    & = & q_{13}\frac{k_x k_y}{k^2} \, ,
\end{eqnarray}
and 
\begin{eqnarray} 
\tilde{{\cal Q}}_{12}({\bf k}) & = & -\frac{a_3 - 2 a_2}{2 a_1+a_3}\,\frac{k_x^2 -  k_y^2}{k^2} \nonumber \\
 & = & -q_{12}\frac{k_x^2 - k_y^2}{k^2}.
\end{eqnarray}
\label{kerdef}
\end{subequations}
In effect, the above equations of constraint, connect the instantaneous $e^A_1$ to the 
dynamical $e^P_1$,  $e_2$ and $e_3$.
\begin{figure}
\begin{center}
\includegraphics[width=4cm]{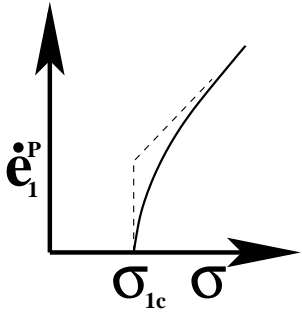}
\end{center}
\caption{Pictorial representation of the dynamics of the plastic strain $e_1^P$as a function of the local volumetric stress $\sigma_1$ given by  
(\ref{plastrain}) for a typical flowing solid (solid line) and the 
simplified form (\ref{micplas}) used in our computations (dashed line).
}
\label{epfig}
\end{figure}

\begin{figure}
\begin{center}
\includegraphics[width=8.5cm]{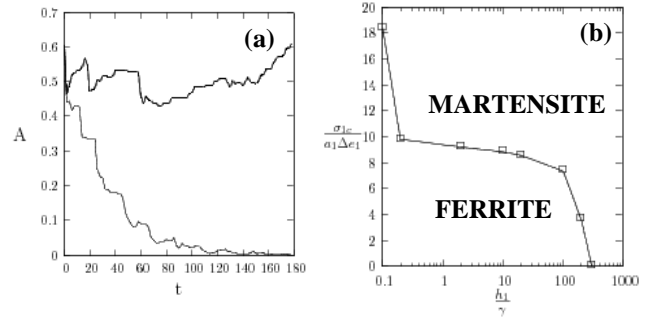}
\end{center}
\caption{Results from the numerical solution of the dynamical equations 
in a $128\times128$ grid with a time step $\delta t = .002$. The parameters
for this calculation are $a_1 = 100, a_2 = 1, a_3 = .01, \gamma = 5$.
(a) Shape asphericity $A$ of the growing nucleus as a function of 
time $t$ from the elastoplastic model  -- bold line martensite and 
thin line ferrite. (b) Dynamical phase diagram in the $\sigma_{1c} - h_1$ plane (expressed
in units of $a_1 \Delta \e^A_1$ and $\gamma$, respectively),
starting from the same (elliptical) initial seed. The  
shape asphericity $A$ at later times and the order
parameter strain $e_3$ have been used to determine the phases.}
\label{dpd}
\end{figure}

We now have to specify the dynamics for the OP strains and the plastic NOP strain. At 
this stage we make the approximation of replacing the value of the OP $e_2$ by its 
value at equilibrium, i.e., $e_2 = 0$ for all times. This greatly simplifies the calculation
without changing the physics. 
The dynamical equation for the affine OP strain (\ref{elastic}) $e_3$ may be derived 
from Newton's laws\cite{strn-only} incorporating 
dissipation via a Rayleigh dissipation functional\cite{landau},
\begin{equation}
\f{\partial^2 e_3}{\partial t^2} = \nabla^2\left[\f{\delta {\cal F}}{\delta e_3}+ \gamma \f{\partial e_3}{\partial t}\right]\, ,
\label{dyn1}
\end{equation}
where $\gamma$ is a solid shear viscosity.  

The dynamics of the plastic NOP strain are determined by (\ref{plastrain}), 
\begin{eqnarray} 
\label{micplas}
\dot {e}_{1}^P & = & \frac{1}{h_1}\,\,\,\sigma_{1}\\ \nonumber
                     &   & \,\,\,\,\,\,\,\,\,\,\,\,\,\,\,\,\,\,\,{\rm if}\, \vert \sigma_{1}\vert  > \sigma_{1c} \\ \nonumber
               & = & 0\,\,\,\,\,\, {\rm otherwise} 
\end{eqnarray}
where we have, for simplicity, chosen $\theta =1$ and a simple threshold criterion, with yield stress $\sigma_{1c}$.

We have solved the dynamical equations (\ref{dyn1}) and (\ref{micplas})
by discretizing in space and time. The initial conditions are chosen from an appropriately
seeded (metastable) square phase (see discussion at the end of previous subsection). Details of the numerical computation are given in Appendix B.  Note all quantities have been made dimensionless.

As in the MD simulation, we first obtain a dynamical phase diagram demarcating the regions
where a martensite or ferrite is obtained upon quenching. 
This is done by studying the shape of the nucleus, in terms of the shape asphericity 
$A$, and twinning of the microstructure, in terms of the affine OP strain $e_3$.
We focus on a special cut in parameter space; we fix the coefficients
appearing in (\ref{frengy}) and $\gamma$, and explore the dynamical phase diagram in the
plasticity variables, $\sigma_{1c}$--$h_1$ plane (Fig.\ref{dpd}(b)).
The threshold stress $\sigma_{1c}$ is expressed in units of $a_1 \Delta \e^A_1$ (the affine stress
at the structural transition), and $h_1$ in units of $\ell^2/\gamma$ (where $\ell$, thickness of the twin
interface, has been taken to be $1$).
In Fig.\ref{dpd}(a), we plot $A$ as a function of time for an anisotropic (twinned) nucleus
and an isotropic (untwinned) nucleus; the value of $A$ at late times (Fig.\ref{dpd}(a)),
and the profile of the order parameter strain is used to map out a dynamical phase diagram 
containing the martensite and ferrite, Fig.\ref{dpd}(b).

We discuss several novel features of this dynamical phase diagram. For instance, even when the threshold stress $\sigma_{1c}$ is zero, a martensite can form if the plasticity relaxation rate is small (large $h_1$) compared to the rate of growth of the nucleus. 
This feature was already present in our earlier calculation\cite{ourprl}, where the dynamics of the local
density fluctuations determined the selection of microstructure, and is an inescapable feature of real martensites\cite{bhadeshia}. The phase diagram, Fig.\ref{dpd}(b), is constructed for a fixed value of
under-cooling. As the degree of under-cooling changes, the phase boundary changes slightly, but not a
whole lot. More significantly, the plasticity relaxation time $h_1$ increases with the lowering of temperature. Thus by starting out in the ferrite phase, one can cross the phase boundary into the
martensite by simply lowering the temperature, identified as the martensite-start or $M_s$ 
temperature. In addition, there is a
well defined plateau yield stress over three decades in $h_1$, suggesting that the
yield stress is independent of temperature over this range. This is consistent with
our MD simulations. 
Finally, it must be noted that the dynamical phase diagram Fig.\ref{dpd}(b) is constructed from the
nature of the first critical nucleus that forms. In a macroscopic sample,
a ferrite nucleus may eventually nucleate and grow
{\it even in the martensite phase}, once the plastic strain $e_1^P$ 
gets enough time to relax. This is consistent with our MD simulations and agrees with results of isothermal quenching 
experiments in real materials\cite{bhad}.

\begin{figure}
\begin{center}
\includegraphics[width=8.5cm]{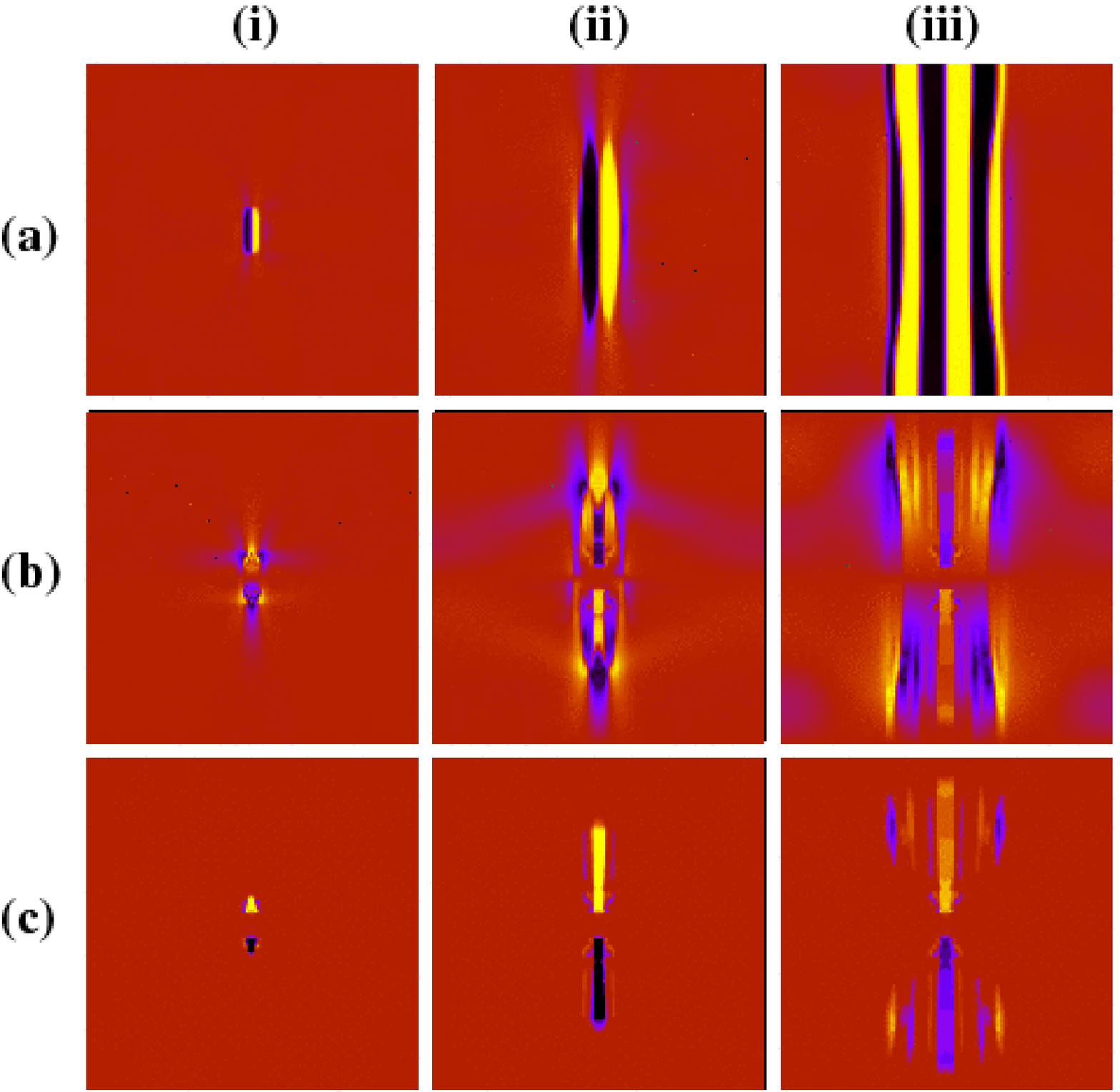}
\end{center}
\caption{(color-online)~Time development of the affine strain, stress and 
non-affine strain following a quench into the martensite phase for 
$256\times256$ cells, at (i) $t = 40$, (ii) $t= 350$ and (iii) $t=800$, 
starting from an initial elliptical nucleus with a single twin boundary. 
The plasticity parameters $|\sigma_{1c}| = 1$ and $h_1 = 1$ while the rest 
of the parameters are as before.
(a) Profile of affine OP strain $e_3$, showing the initial growth parallel to 
the twin interface, followed by the dynamical addition of twins. Colors: 
yellow to black maps the 
range $-1. < e_3 < 1$. Brown region denotes retained austenite, $e_3 = 0$. 
(b) Corresponding profile of the local stress 
$\sigma_1$. The local stress is concentrated at the tips of the growing 
front where it approaches the threshold value $\sigma_{1c}$ (i) and (ii). 
In the interior of the growing nucleus $\sigma_{1}$ relaxes to zero.  
Subsequently in (iii), the local stress gets large in regions where the 
new twins are being accommodated. Note the variation in the signs of the 
stress in the direction along which new twins are
added. Colors: yellow to black maps the range $-1. < \sigma_{1} < 1$. 
(c) Corresponding non-affine strain $e_1^P$, showing the initial advection by 
the transformation
front, and its dynamical emergence as subsequent twins are added. $e_1^P$ appears in regions
where $\sigma_1 \sim \sigma_{1c}$ and $\sigma_1 \sim - \sigma_{1c}$.
Colors: yellow to black maps the range $-.01 < e_1^P < .01$.}
\label{mart}
\end{figure}
\begin{figure}
\begin{center}
\includegraphics[width=6.cm]{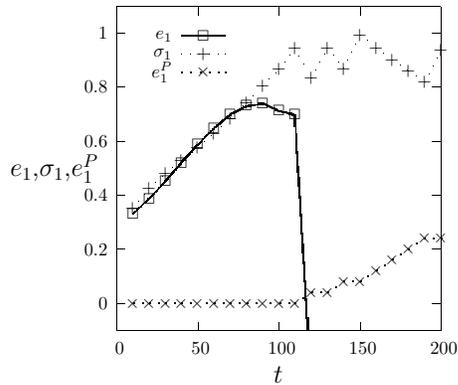}
\end{center}
\caption{Evolution of the local total strain $e_1$, the local 
stress $\sigma_1$ and the non-affine part of the strain $e_1^P$ for
a cell $\Omega = (128,100)$ on the twin axis following the quench into 
the martensite phase shown in Fig.\ref{mart}. Note the initial linear
regime, when $\sigma_1 \propto e_1$ and $e_1^P = 0$, followed by 
oscillations in $\sigma_1$ and the creation of $e_1^P$ as the stress 
$\sigma_1$ rises to the threshold value $\sigma_{1c} = 1$.  
The resemblance with Fig.\ref{timeld} from our MD simulations is quite 
apparent. We have multiplied the strains by $100$ in order to plot them to the 
same scale.}
\label{timev}
\end{figure}
\begin{figure}
\begin{center}
\includegraphics[width=8.5cm]{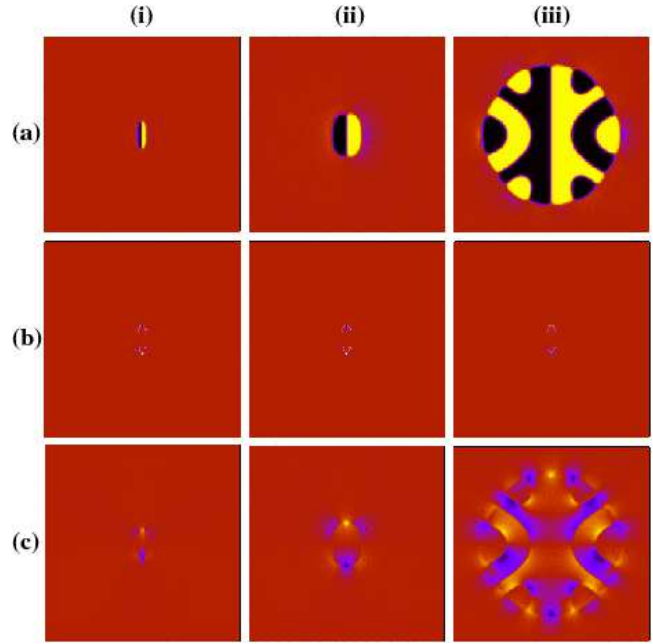}
\end{center}
\caption{(color-online)~
Time development of the affine strain, stress and non-affine strain following 
a quench into the ferrite phase, at (i) $t = 40$, (ii) $t= 300$
and (iii) $t=1600$, starting from an initial elliptical nucleus with a 
single twin boundary. 
The plasticity parameters $\sigma_{1c} = 0$ and $h_1 = .1$ while the rest
of the parameters are as before.
(a) Profile of affine OP strain $e_3$, which shows the initial elliptical nucleus growing approximately isotropically. Colors: yellow to black maps the 
range $-1. < e_3 < 1$. (b) The local stress $\sigma_1$ rapidly relaxes to 
zero, in this case, with only a small residual value remaining in the vicinity 
of the original seed. Colors:yellow to black maps the range 
$-1. < \sigma_1 < 1$. 
(c) Corresponding non-affine strain $e_1^P$, showing its 
invasion into the `bulk' of the growing polycrystalline nucleus. 
Colors: yellow to black maps the range $-.1 < e_1^P < .1$.}
\label{fer}
\end{figure}
Having displayed the dynamical phase diagram we 
can perform quenches to the martensite and ferrite phase, and study the time development of the 
profiles of $e_3$, $\sigma_1$ and $e^P_1$ (Figs.~\ref{mart}, \ref{fer}).
Using plasticity parameters corresponding to the martensite phase, 
(Fig.~\ref{mart}) shows the 
temporal evolution of a twinned nucleus, in perfect analogy 
with our MD simulations. The nucleus initially grows parallel
to the twin boundary (Fig.~\ref{mart}(i)), while the stress $\sigma_1$ 
approaches the threshold at the growing tips. As a result the plastic strain 
$e^P_1$ gets to be large at these tips. As the nucleus grows, these highly 
stressed and plastic regions are advected by the growing tips. In the interior 
transformed region, the stress relaxes to a sub-threshold value, and the 
plastic deformation goes to zero. The sequence of events exactly mimic the 
dynamics of the $\e^A_V$, $\chi$ and NAZs of Fig.~\ref{martsim}. 
To study the time evolution of the NAZs in  more detail, we focus on 
a single cell, $\Omega$, midway within our computation box, along 
(but not on) the twin boundary of the initial seed within the 
untransformed square lattice. With time, the growing tip of the nucleus 
approaches, and then sweeps by $\Omega$, in the process transforming it into 
the triangular phase. This situation is analogous to that shown in 
Fig.\ref{timeld} from our MD simulations. We plot the local $e_1$,$\sigma_1$ 
and $e_1^P$ at $\Omega$ as a function of time $t$ in 
Fig.\ref{timev}. As in Fig.\ref{timeld}, initially $\Omega$, which lies
ahead of the approaching transformation front, begins to deform elastically 
due to stress generated at the growing tip. The resulting volumetric strain 
$e_1$ is proportional
to the local $\sigma_1$ and $e_1^P = 0$. As the tip of the growing nucleus 
approaches $\Omega$, $\sigma_1$ rises and tends to cross the threshold, 
$\sigma_{1c}$. At this instant, $e_1^P$ begins to form reducing $\sigma_1$
to a value below $\sigma_{1c}$. As the nucleus grows further, $\sigma_1$ 
within $\Omega$ increases again -- and the process repeats producing a local 
stress which oscillates rapidly in time. 
These oscillations result from cooperative jamming and unjamming events caused 
by alternating build-up of $\sigma_1$ due to interface motion and its 
relaxation by creation of NAZs\cite{liunagel}. The region of high 
(and oscillating) local stress and the NAZ travels with the growing 
tip, being advected by the moving transformation front. The strong resemblance 
between Figs.\ref{timeld} and \ref{timev} is striking. Eventually, the 
interface crosses $\Omega$ and $\sigma_1$, as well as $e_1^P$ relaxes to 
zero within the bulk of the product phase. 

The subsequent dynamics, Fig.\ref{mart}(iii), goes beyond the time scales accessed in the MD  simulation. The dynamics now proceeds perpendicular to the twin interface, adding 
new twins (symmetrically disposed) as time proceeds. 
The production of 
new twins with a fixed width, is a consequence of the anisotropic, non-local interactions connecting spatially 
separated regions with nonzero $e_3$. 

Note that the affine NOP strain mediates the anisotropic, long-ranged interactions\cite{strn-only}; the presence of the plastic NOP strain {\it screens} this interaction, making it short-range (Appendix B). The emergence of the OP strain in the form of twins, leads to 
an increase in the local stress $\sigma_1$, which in turn generates plastic flow on crossing the threshold (Fig.\ref{mart}(c)). The plastic strain, once 
produced, reduces the value of the total NOP strain and, therefore, that of 
the non-local interaction. In the case of the martensite, 
this reduction is not complete. The stress $\sigma_1$ decreases to zero in the interior of the martensitic nucleus, so that any given region undergoes the same sequence of transformations : 
untransformed $\to$ elastic distortion $\to$ non-affine $\to$ transformed. This temporal sequence is also seen in our MD simulations.

We now use plasticity parameters corresponding to the ferrite phase; Fig.\ref{fer} depicts the time  evolution of an isotropic, polycrystalline ferrite nucleus, starting from the same initial conditions
as above. For small values of $\sigma_{1c}$, the plastic strain $e_1^P$ is produced readily, and on an average largely cancels out the effect of the affine 
NOP strain. This significantly reduces the magnitude and range of the non-local 
interactions, which were responsible for producing the twins. The local stress 
$\sigma_1$ tends to cross the threshold (positive and negative) in the 
{\it interior} of the growing nucleus, which leads to 
an {\it invasion} of the non-affine strain $e_1^P$ into the `bulk' of the 
isotropically growing, polycrystalline nucleus. This results in incoherent 
grain boundaries in the interior of the growing nucleus. Note that the 
symmetry of the pattern, Fig.\ref{fer}, arises because the evolution equations 
are deterministic; any noise would destroy this symmetry and 
make the grain boundaries rough and orient randomly.
The sequence of events then exactly mimic the dynamics of the $\e^A_V$, 
$\chi$ and NAZs of Fig.~\ref{ferritesim}. 

We believe we have successfully constructed a general elastoplastic description for the dynamics of
solid state transformations, which is capable of describing different microstructures and addresses the issues of microstructure selection. The qualitative picture that emerges from the
elastoplastic model closely resembles our MD simulation results. In the next section, we will present some implications of our elastoplastic theory, vis-a-vis the discussion in the Introduction.

\section{Discussion}
By constructing a theory for microstructure selection, we successfully bridge two apparently disparate
descriptions of the dynamics of ferrites and martensites, within a unified framework. This has been
achieved at the cost of enlarging the space of dynamical variables to include non-affine deformations.
Have we lost some of the special features of martensites in the process ? What implications does our 
 description
of solid state nucleation and microstructure selection have for conventional nucleation theory?

\begin{enumerate}

\item {\it Emergence of average compatibility from the elastoplastic dynamics} : Our elastoplastic theory
 provides an understanding of how the strict local compatibility of strain-only theories\cite{strn-only,rajiv} can be reconciled with the average compatibility of geometrical theories\cite{geom}, when describing the dynamics of the martensite. We find that in the regime where the martensite obtains, our dynamical equations reduce to the equations of the strain-only theories, when we coarse-grain over a scale $\lambda$ corresponding to the size of the NAZs. Simultaneously, the equations of constraint, viz.,
 the modified St. Venant's condition (\ref{stven}), (\ref{venant}), reduces to the usual 
St. Venant's elastic compatibility, provided we coarse-grain over the same scale $\lambda$.
This can be seen by explicitly writing out (\ref{venant}),
$$  
{\bf \nabla}^2 e^A_1 - (\nabla_x^2 - \nabla_y^2)e_2 - 4\nabla_x \nabla_y e_3 = 
{\bf \nabla}^2 e_1^P.
$$ 
In Fig.\ref{corse}, we have re-plotted $e_1^P$ for 
time $t=800$ (Fig.\ref{mart}(c)(iii)) in order to show the NAZs in detail.
Remarkably, the martensite plates are accompanied by patches
where $e_1^P \neq 0$ and alternate in sign. If the system is coarse-grained 
over a distance $\lambda$ the effect of these patches cancel and we recover 
the usual elastic compatibility condition. This emergence of compatibility upon coarse-graining over the scale of the plastic zone was noted in \cite{drop}. As the critical stress $\sigma_{1c}$ increases, 
the size of the NAZs decreases and so does the coarse-graining 
length scale $\lambda$. 
Note that the coarse-graining appropriate for the emergence of the usual elastic compatibility, does not wash out the twinned microstructure of the martensite; the `phase' of the averaging is so as to produce a planar martensite-austenite interface as 
shown by the dashed line in Fig.\ref{schematic}.
\begin{figure}
\begin{center}
\includegraphics[width=5.0cm]{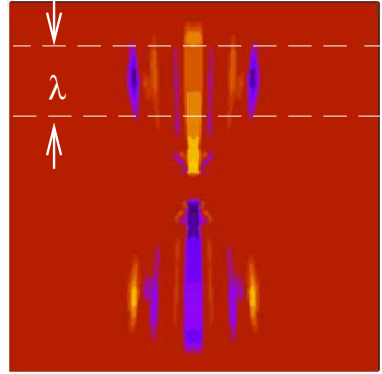}
\end{center}
\caption{(color-online)~ The plot of the plastic strain $e_1^P$ for a 
growing martensite at $t = 800$ (same as in Fig.\ref{mart}(c)(iii)). 
Note the presence of NAZs with alternating signs of $e_1^P$ accompanying the 
growing martensite phase -- see region on top the within the dashed lines,  
a similar region (not marked) exists near the bottom of the figure. This 
figure needs to be compared with Fig.\ref{schematic}.
When the strain fields are coarse-grained over the length scale $\lambda$ 
the contribution from these alternating patches cancel, and full elastic 
compatibility is restored.}
\label{corse}
\end{figure}

\item {\it Reversibility of martensitic transformations} : How do we reconcile the inevitable creation and evolution of plasticity in the form of NAZs with 
the apparent reversibility of martensitic transformations, as observed in shape-memory alloys ? 
Here we will provide some preliminary comments, which will be taken up in 
greater detail later\cite{reversibility}. In essence, microstructural 
reversibility in martensites, is related to the nature of the accompanying 
plastic deformation. The key feature of plastic deformation in the NAZs, is 
that it is largely associated with the NOP sector, which in turn is 
{\it slaved} to the transformation strain. Indeed even the slightest amount 
of plasticity in the OP sector, would make the transformation 
irreversible. This is apparent in martensites involving Fe 
alloys, which do not exhibit shape-memory\cite{kaunat}. Within our own model 
system, a deep quench to the $\alpha = 0$, $v_3 = 0$ region produces a 
{\em triangular} solid which is not related to the parent square lattice by a 
group-subgroup relation\cite{Avad-sym}. During the reverse transformation, therefore,
there is no unique parent lattice that the system can revert to. This produces 
non-affineness in the OP sector due to a multiplicity of affine paths and destroys 
reversibility. Our results related to such non-reversible transformations will 
be published elsewhere. 

The other relevant feature exhibited by the NAZs associated with martensites, is the special 
nature of the particle trajectories\cite{ourprl,jayeetraj}. Particles in the NAZs formed 
during martensitic growth, move ballistically and in a coordinated manner. It is these two 
properties of the NAZs discussed here that ultimately renders the square to rhombus 
martensitic transformation reversible, in spite of significant transient and localized 
plastic deformation.

\item {\it Inconsistency with Ostwald's step rule} : Ostwald's step rule of 1897, states
that {\em ``the phase that nucleates need not be the stable phase, but the one that is
closest in free energy to the parent phase \ldots''}. This rule has been interpreted
by Stranski and Totomanov\cite{ostwald} to mean that the phase which has the 
lowest free-energy barrier is nucleated.
While it is easy to appreciate the applicability 
of this rule for phase transformations in simple systems having a uniquely defined barrier 
crossing event along the path of the transformation, it is 
more difficult to apply such considerations to, say, atomic rearrangements and 
the generation of NAZs where many barriers with different attempt frequencies
may be involved\cite{osy4}.   As discussed in section III, the selection
of microstructure depends both on parameters in the free-energy functional (\ref{frengy})
and dynamical parameters in (\ref{dyn1}) and (\ref{micplas}). This is explicitly shown
in Fig. \ref{dpd}, where the microstructure depends on the plasticity dynamics, in terms of the yield stress $\sigma_{Vc}$ and viscosity $h$. 

\item {\it Randomness and heterogeneous nucleation at defect sites} : 
The results presented in section III A, were obtained with initial conditions corresponding to a 
{\it small elliptical nucleus} and the choice of dynamical parameters corresponding to martensitic and ferritic
growth. An alternate initial condition for the nucleation dynamics is to prescribe a spatially random
stress profile, e.g., a random $\sigma_1^{\prime}$. In real materials, this would correspond to frozen in defect structures. This quenched random stress would add to the internal
stress so that the total stress $\sigma_1 = a_1 e_1^0 + \sigma_1^{\prime}$.
The {\it intial} stages of the dynamics of nucleation and growth is sensitive to the initial distribution, which we
take from a Gaussian distribution with zero mean
and width $S_1$.

For instance, if $S_1 \ll \sigma_{1c}$, the threshold, inhomogeneous elastic strains in $e_1$ develop which initiate nucleation of the product, via its coupling to $e_3$. As the dynamics proceeds, local (total) $\sigma_1$ gets enhanced at the transformation fronts, giving rise to plastic deformations.
On the other hand, if $S_1 > \sigma_{1c}$, the threshold, then local regions can develop appreciable plastic strains, Eqn.\ref{micplas}. The elastic coupling between $e_1^P$ and $e_3$ through
the functional derivative in Eqn~\ref{dyn1} then causes the nucleation of
the product phase (see Appendix B). The lag time for nucleation is appreciably faster compared to when $S_1 \ll \sigma_{1c}$. The qualitative features of the  phase diagram, Fig.\ref{noisy}, are unaltered by the presence of quenched random stress fields, though it `enlarges' the regime over which ferrite phase obtains. The ferrite microstructure is altered; the grain sizes shrink and the grain boundaries thicken with increased randomness. 
The martensite microstructure, at the scale of the twin pattern is however unaltered, though again the 
the distribution of martensitic grains gets smaller with increased disorder. This robustness of martensitic patterning over the scale of the twins is significant, and we wish to revisit this aspect in a detailed study.

\begin{figure}
\begin{center}
\includegraphics[width=8.8cm]{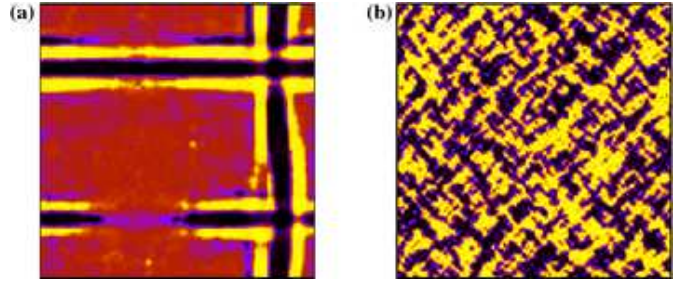}
\end{center}
\caption{(color-online)~Plot of the OP strain $e_3$ for time
$t = 1000$ for a Martensite (a) with $\sigma_{1c}/a_1 \Delta e_1 = 2500$ and 
for a ferrite (b) with $\sigma_{1c} = 0$; $h_1/\gamma = .1$ for (a) and (b).
The color scheme is as in Fig.\ref{mart}(a).  
In both these calculation, a quenched, Gaussian, random 
stress $\sigma_1({\bf r},t)$ with zero mean and variance $S_1 = 10$ was 
used to nucleate the product phase. Note that quenched randomness 
does not affect the overall characteristics of 
the Martensite. The ferrite grains however become much smaller with grain 
boundaries which are broad compared to the size of the grains.  
}
\label{noisy}
\end{figure}

\item {\it Dynamical phase diagram and TTT curves} : The dynamics of microstructure selection is conventionally represented by a Time-Temperature-Transformation (TTT) diagram, constructed
 in the form of contour plots of the 
proportion of each constituent (martensite or ferrite) as a function of time during an isothermal 
transformation at different quench temperatures\cite{bhad,hasen}. Figure \ref{dpd}, is the 
dynamical phase diagram computed within our elastoplastic theory; this can be  
converted to the typical TTT curve, provided
we know the temperature dependence of $\sigma_{1c}$ and $h_1$. Such a temperature dependence
may be put in phenomenologically, as in Fig. 4 of \cite{ourprl}, or obtained from a first principles
non-affine field theory.

\item {\it Future work} : It should be possible to extend our elastoplastic theory to include the
effects of impurities such as interstitial carbon in Fe (as in steel)\cite{rttt}, which undergoes
significant non-affine deformation\cite{bhad}. Interstitial carbon
represented by a diffusive concentration field $\psi$, would enter into both the dynamical
equations for the affine and non-affine strains. Thus any attempt to understand microstructure
selection in systems such as steel, would involve the study of the 
coupled dynamics of the affine strain, non-affine strain and concentration field $\psi$.

\end{enumerate}

 Finally, apart from restoring the full
tensorial character of the elastoplastic description, we need to explore in greater detail, the consequences of general 
thresholding and yield flow in the plasticity dynamics accompanying solid state transformations.
In a later paper\cite{reversibility}, we will discuss the dynamical response of the transforming solid to
time dependent external stresses (or strains), and periodic quenches across the phase boundary.

\acknowledgements
We acknowledge a series of illuminating conversations and correspondences with 
K. Bhattacharya and A. Saxena, and their useful comments on the manuscript. We also 
thank G.I. Menon for a careful reading. Computer time from DST grant SP/S2/M-20/2001 
and support from the Unit for Nano Science and Technology, S. N. Bose 
National Centre for Basic Sciences is gratefully acknowledged. 

\appendix
\section{}
In this appendix, we reproduce the formul\ae~ used to calculate the energy,
forces and stresses for the three-body potential used. The form of the 
three-body potential ensures that these quantities are 
computable using pairwise functions alone\cite{still}. This greatly speeds 
up the computation.\\

\noindent
{\it Energy} : The three-body part of the energy is given by
\begin{eqnarray}
E_3 &=& \frac{1}{6}\sum_{i\ne j\ne k} \psi_3(r_{ij},r_{jk},r_{ki}) \\ \nonumber
    &=& \frac{1}{2}\sum_{i\ne j\ne k} 
\frac{V_3}{4}f_{ij} \sin^2(4\theta_{jik})f_{ik} \\ \nonumber 
     &=& \sum_{i\ne j\ne k} 2\left(\sin^2(\theta_{jik}) \cos^2(\theta_{jik})\right. \\ \nonumber
&-&\left. 4\sin^4(\theta_{jik}) \cos^4(\theta_{jik})\right)f_{ij}f_{ik}. 
\end{eqnarray}
Now define $\tilde{x}_{ij} = x_{ij}/r_{ij}$ and $\tilde{y}_{ij}=y_{ij}/r_{ij}$,
 so that $\sin\theta_{jik} = \tilde{x}_{ik}\tilde{y}_{ij} - 
\tilde{x}_{ij}\tilde{y}_{ik}$ and $\cos\theta_{jik} = \tilde{x}_{ij}\tilde{x}_{ik} + 
\tilde{y}_{ij}\tilde{y}_{ik}$. Using the above definitions and the quantities 
\begin{eqnarray*}
g_{ij}(1) &=& \tilde{x}_{ij}^2\tilde{y}_{ij}^2f_{ij}, \\
g_{ij}(2) &=& \tilde{x}_{ij}^2\tilde{y}_{ij}^2 
(\tilde{x}_{ij}^2 - \tilde{y}_{ij}^2)f_{ij},  \\
g_{ij}(3) &=& \tilde{x}_{ij}^4\tilde{y}_{ij}^4f_{ij}, \\
g_{ij}(4) &=& \tilde{x}_{ij}^2\tilde{y}_{ij}^2
(\tilde{x}_{ij}^2 - \tilde{y}_{ij}^2)^2f_{ij}, \\
g_{ij}(5) &=& \tilde{x}_{ij}^3\tilde{y}_{ij}^3
(\tilde{x}_{ij}^2 - \tilde{y}_{ij}^2)f_{ij}, 
\end{eqnarray*}
we get $E_3 = V_3\sum_{i} S_i$ with 
\begin{eqnarray*}
S_i &=& 4\left[G_i(1)F_i - 4G_i(1)^2 - G_i(2)^2\right] \\ \nonumber
&-& 16\left[G_i(3)F_i + 32G_i(3)^2 + 2G_i(4)^2 \right. \\ \nonumber
&+&\left. G_i(1)^2 - 16G_i(3)G_i(1) \right. \\ \nonumber
&-&\left. 4G_i(5)G_i(2) + 16G_i(5)^2 \right]
\end{eqnarray*}
where $G_i(n) = \sum_{j\ne i} g_{ij}(n)$ and $F_i =  \sum_{j\ne i} f_{ij}$.\\

\noindent
{\it Force}: The three-body forces can be found by taking derivatives of $E_3$ which can be
 cast into similar forms. Remembering that $\tilde{y}_{ij}$ implicitly depends 
on $x_{ij}$ through $r_{ij}$ and evaluating the quantities 
\begin{eqnarray*}
\frac{\partial g_{ij}(1)}{\partial x_{ij}} &=& 
2\left[\tilde{x}_{ij}\tilde{y}_{ij}^2(1-\tilde{x}_{ij}^2) \right. \\ \nonumber
&-&\left.\tilde{x}_{ij}^3\tilde{y}_{ij}^2\right]f_{ij}/r_{ij}\\ \nonumber 
&+& 2\tilde{x}_{ij}^3\tilde{y}_{ij}^2(r_{ij}-r_0), \\
\frac{\partial g_{ij}(2)}{\partial x_{ij}} &=& 
\left[(3\tilde{x}_{ij}^2\tilde{y}_{ij}-\tilde{y}_{ij}^3)
(1-\tilde{x}_{ij}^2) \right. \\ \nonumber
&-&\left. \tilde{x}_{ij}\tilde{y}_{ij}(\tilde{x}_{ij}^3 - 
3\tilde{x}_{ij}^2\tilde{y}_{ij}^3)\right]f_{ij}/r_{ij}  \\ \nonumber 
&+& 2\tilde{x}_{ij}^2\tilde{y}_{ij}(\tilde{x}_{ij}^2 - 
\tilde{y}_{ij}^2)(r_{ij}-r_0),\\
\frac{\partial g_{ij}(3)}{\partial x_{ij}} &=& 
\left[4\tilde{x}_{ij}^3\tilde{y}_{ij}^4(1-\tilde{x}_{ij}^2) - 
4\tilde{x}_{ij}^5\tilde{y}_{ij}^4\right]f_{ij}/r_{ij} \\  \nonumber
&+& 2\tilde{x}_{ij}^5\tilde{y}_{ij}^4(r_{ij}-r_0),\\
\frac{\partial g_{ij}(4)}{\partial x_{ij}} &=& 
\left[(6\tilde{x}_{ij}^5\tilde{y}_{ij}^2
 + 2\tilde{x}_{ij}\tilde{y}_{ij}^6 \right. \\ \nonumber
&-&\left. 8\tilde{x}_{ij}^3\tilde{y}_{ij}^4)(1-\tilde{x}_{ij}^2)
-(6\tilde{y}_{ij}^5\tilde{x}_{ij}^2 \right. \\ \nonumber
&+&\left. 2\tilde{y}_{ij}\tilde{x}_{ij}^6
 - 8\tilde{y}_{ij}^3\tilde{x}_{ij}^4)\tilde{x}_{ij}\tilde{y}_{ij}
\right]f_{ij}/r_{ij} \\  \nonumber
&+& 2\tilde{x}_{ij}^3\tilde{y}_{ij}^2(\tilde{x}_{ij}^2 
- \tilde{y}_{ij}^2)^2(r_{ij}-r_0),\\
\frac{\partial g_{ij}(5)}{\partial x_{ij}} &=& 
\left[(5\tilde{x}_{ij}^4\tilde{y}_{ij}^3 - 3\tilde{x}_{ij}^2\tilde{y}_{ij}^5)
(1-\tilde{x}_{ij}^2) \right. \\ \nonumber
&-&\left. (3\tilde{x}_{ij}^5\tilde{y}_{ij}^2
- 5\tilde{x}_{ij}^3\tilde{y}_{ij}^4)\right]f_{ij}/r_{ij} \\ \nonumber
&+& 2\tilde{x}_{ij}^4\tilde{y}_{ij}^3(\tilde{x}_{ij}^2 - 
\tilde{y}_{ij}^2)(r_{ij}-r_0),
\end{eqnarray*}
the force acting on the particle 'i' in the $x$-direction owing to the 
three-body interaction can be written as
\begin{eqnarray}
F^x_i = \sum_{j\ne i}{\it H^x_{ij}}
\end{eqnarray} 
where
\begin{eqnarray*}
{\it H^x_{ij}}& =& 4V_3\left[\frac{\partial g_{ij}(1)}{\partial x_{ij}}
\sum_{k\ne i}f_{ik} + 2\tilde{x}_{ij}(r_{ij}-r_0)\sum_{k\ne i}g_{ik}(1)\right. 
\\ \nonumber
&-&\left. 8\frac{\partial g_{ij}(1)}{\partial x_{ij}}\sum_{k\ne i}g_{ik}(1)
- 2\frac{\partial g_{ij}(2)}{\partial x_{ij}}
\sum_{k\ne i}g_{ik}(2)\right] \\ \nonumber
&-&16V_3\left[\frac{\partial g_{ij}(3)}{\partial x_{ij}}\sum_{k\ne i}f_{ik}
+2\tilde{x}_{ij}(r_{ij}-r_0)\sum_{k\ne i}g_{ik}(3) \right.\\ \nonumber 
&+&\left.64\frac{\partial g_{ij}(3)}{\partial x_{ij}}\sum_{k\ne i}g_{ik}(3) + 
4\frac{\partial g_{ij}(4)}{\partial x_{ij}}\sum_{k\ne i}g_{ik}(4)\right. \\ \nonumber
&+&\left.2\frac{\partial g_{ij}(1)}{\partial x_{ij}}\sum_{k\ne i}g_{ik}(1) - 
16\frac{\partial g_{ij}(1)}{\partial x_{ij}}\sum_{k\ne i}g_{ik}(3) \right. \\ \nonumber
&-&\left.4\frac{\partial g_{ij}(5)}{\partial x_{ij}}\sum_{k\ne i}g_{ik}(2) 
- 4\frac{\partial g_{ij}(2)}{\partial x_{ij}}\sum_{k\ne i}g_{ik}(5) \right. \\ \nonumber
&+&\left. 32\frac{\partial g_{ij}(5)}{\partial x_{ij}}\sum_{k\ne i}g_{ik}(5)\right].
\end{eqnarray*}
The $y$-component of the force 
\begin{eqnarray} 
F^y_i = \sum_{j\ne i} {\it H^y_{ij}},
\end{eqnarray} 
where $H^y_{ij}$ is evaluated using expressions similar to that for $H^x_{ij}$
given above. \\

\noindent
{\it Stress} : The contribution of the three-body interaction to the virial stress can now
be calculated using the force components,
\begin{equation}
\sigma_{\alpha \beta} = \frac{1}{2}\sum_{i,j} r^{\alpha}_{ij}{\it H^{\beta}_{ij}}
\end{equation}
where $\alpha, \beta, = x,y$.

\section{}

In this appendix we give details of the numerical solution of (\ref{dyn1})
and (\ref{micplas}) used to obtain the results of section III. We have used a 
simple real space scheme for the solution, discretizing the partial
differential equations over a lattice of square cells of size $\delta x = 1$.
We have used $128\times128$ cells to obtain the phase diagram (Fig.\ref{dpd})
and $256\times256$ cells for Figs.\ref{mart},\ref{timev} and \ref{fer}.  
The initial value problem in time is solved using an Euler scheme with 
a time step of $\delta t = .002$ which is sufficient to avoid numerical 
instabilities. Below we give the sequence of steps involved in the iteration 
of the discretized equations.  

\vskip .1cm
\noindent
{\em Step 1.} We start with initial values for $e_3$, its time derivative 
$\dot e_3$ and $e_1^P$ defined over all the cells in our lattice and for 
time $t$. First, we need to compute the (slaved) affine strain $\e_1^A$ from 
the OP strain $e_3$ by solving (\ref{kerdef}a). The affine strain $e_1^A$ 
together with the known $e_1^P$ determines the total strain $e_1$ at $t$. 
In real space, \ref{kerdef}a  becomes,
\begin{equation}
\nabla^2\,e_1(x,y) = q_{13} \frac{\partial^2}{\partial x \partial y} e_3(x,y)
\label{real13}
\end{equation}
Equation \ref{real13} is the Poisson equation 
for the charge density,
\begin{equation}
\rho_3(x,y,\{e_3\}) = q_{13} \frac{\partial^2}{\partial x \partial y} e_3(x,y)
\end{equation}
and we need to solve it for the (Dirichlet) boundary condition $e_1 \to 0$ 
for $x,y \to \infty$. This is done by discretization in 
real space and by using an iterative scheme with a small over-relaxation and 
a convergence criterion of $1$ in $10^{6}$\cite{abram}. 
For convenience in what follows, we refer to this solution using the notation,
$e_1 = {\cal P}(\{\rho_3\})$. 

Our numerics can be checked for accuracy by comparing the results with that 
of simple choices for $e_3$ for which $e_1^A$ (same as $e_1$) 
may be obtained analytically.
For example for $e_3$ which is nonzero only within a square of size $2a$,
viz.,
\begin{equation}
e_3(x,y) = e_0\Theta(a+x)\Theta(a-x)\Theta(a+y)\Theta(a-y), 
\label{ansatz}
\end{equation}
($\Theta(x)$ is
the Heaviside step function) $e_1^A$ is the 
electrostatic potential for a set of four charges $+e_0, -e_0, +e_0$ and 
$-e_0$ at the vertices of a square $(a,a), (a,-a), (-a,-a), (-a,a)$ in 
two dimensions. This is given by, 
\begin{eqnarray}
e_1^A(x,y) & = & \frac{e_0}{2}\left[\ln\left(\frac{[(x-a)^2+(y-a)^2]}{[(x+a)^2+(y-a)^2]}\right.\right. \nonumber \\
           & & \times \left.\left.\frac{[(x+a)^2+(y+a)^2]}{[(x-a)^2+(y+a)^2]}\right)\right].
\label{charge}
\end{eqnarray}
On the interfaces $y = \pm a$ and $x = \pm a$ it is easy to see that 
$e_1^A \sim \pm x $ (and $\pm y$ respectively) except near the corners where 
there are weak logarithmic singularities. Incidentally, this linear 
approximation for $e_1^A$ 
is the same as the initial value of the density fluctuation
$\phi({\bf r},0)$ used in Refs.\cite{drop} and \cite{ourprl}. The presence of 
$e_1^A$ modifies the interfacial energy of a rectangular nucleus of martensite 
of length $L$ and width $W$ containing $N$ twins leading to the experimentally 
observed scaling law $L/N \sim W^{1/2}$\cite{drop}.     

Our numerical result for $e_1^A$ for the choice of $e_3$ given in 
(\ref{ansatz}) reproduces the analytic form (\ref{charge}) to within a few 
percent.  

\vskip .1cm
\noindent
{\em Step 2.} Knowing the strain $e_1$ at time $t$, we next 
update the plastic strain $e_1^P$ to the next time step $t + \delta t$ by 
iterating (\ref{micplas}). For this, 
the local stress $\sigma_1 = a_1 e_1$ (plus any external stress if present) 
is obtained for all the cells and is then used as input to (\ref{micplas}). 

\vskip .1cm
\noindent
{\em Step 3.} Lastly, we have to update $e_3$ and $\dot e_3$ for which 
one needs to compute the functional derivative,
\begin{equation}
\frac{\delta {\cal F}}{\delta e_3} = \frac{\delta {\cal F}_3}{\delta e_3} + \frac{\delta {\cal F}_1}{\delta e_3}
\end{equation}
where the two terms on the right hand side represent functional derivatives of 
the parts of the free energy (\ref{frengy}) involving only $e_3$ and 
$e_1$ respectively. The first term is straight forward and is given by,
\begin{equation}
\frac{\delta {\cal F}_3}{\delta e_3} = -\nabla^2 e_3 + e_3 - 4 (e_3)^3 + 6 (e_3)^5 
\label{nlk1}
\end{equation}
and the second term, after some algebra can be shown to be,
\begin{subequations}
\begin{equation}
\frac{\delta {\cal F}_1}{\delta e_3} = a_1 {\cal P}(\{\rho_1\}) - c_1 \rho_1
\end{equation}
{\rm  where,}
\begin{equation} 
\rho_1(x,y,\{e_3\}) = q_{13} \frac{\partial^2}{\partial x \partial y} e_1(x,y).
\end{equation}
\label{nlk}
\end{subequations}
Note that (\ref{nlk}) involves the total $e_1$ which includes both affine as 
well as the non-affine strain. Even if $e_3 = 0$ to begin with, at subsequent 
times $e_3$ may be created due to the presence of nonzero $e_1^P$. 
One encounters such a situation during the heterogeneous nucleation of 
martensite near defect sites\cite{hasen,stuck} with pre-existing $e_1^P$. This
is similar to the seeding of the martensite nucleus with a single point-vacancy 
as in our MD simulations. Secondly, the resulting 
form for the functional derivatives are highly non-local since they involve 
repeated solutions of the Poisson equation. However, if $e_1^P$ is large 
the total NOP strain $e_1$ vanishes and the non-local coupling 
between spatially separated regions of the OP $e_3$ disappears.


\begin{thebibliography}{99}

\bibitem{hasen}
R. W. Cahn and J. Haasen, {\it Physical Metallurgy}, 4$^{th}$ Edition,
(Elsevier, Amsterdam, 1996);
A.\ G.\ Kachaturyan, {\it
Theory of Structural Transformations in Solids}, (Wiley, NY, 1983).

\bibitem{stuck}
R. Phillips, {\em Crystals, Defects and Microstructures: Modeling Across Scales},
(Cambridge University Press, Cambridge, 2001).

\bibitem{cahn}
J. W. Cahn and J. E. Hilliard, J. Chem. Phys {\bf 28}, 258 (1957);  

\bibitem{drop}
M. Rao and S. Sengupta, \prl {\bf 78}, 2168 (1997);
M. Rao, M. and S. Sengupta, Curr. Sc. {\bf 77}, 382 (1999);
S. Sengupta and M. Rao, Physica (Amsterdam) {\bf 318A}, 251 (2003).

\bibitem{jpcm}
M. Rao and S. Sengupta, J. Phys: Condens. Mat. {\bf 16}, 7733 (2004).

\bibitem{ourprl}
M. Rao and S. Sengupta, \prl {\bf 91}, 045502 (2003).

\bibitem{ums}
D. Frenkel and B. Smit, {\it Understanding Molecular Simulations}, 2$^{\rm nd}$
Edition, (Academic Press, California, 2002) 

\bibitem{bhad}H.K.D.H. Bhadeshia, {\it Bainite in steels}, (Institute of 
Materials, London, 1992). 

\bibitem{paltwin}A. Heymann, A. Stipp, C. Sinn and T. Palberg, J. Coll., Int. 
Sc. {\bf 207}, 119127 (1998); J. Liu and T. Palberg, Progr. Colloid Polym. Sci.
{\bf 123}, 222226 (2004).

\bibitem{geom}
J.M. Ball and R.D. James.  Phil. Trans. Roy. Soc. London A, {\bf 338}, 
389, 1992; K. Bhattacharyya ({\em unpublished notes}) available from 
http://mechmat.caltech.edu/
{\em Martensite} eds. G.\ B.\ Olson and W.\ S.\ Owen, (ASM International,
The Materials Information Society, 1992). 

\bibitem{kaubook}
K. Bhattacharyya, {\em Microstructure of martensite} (Oxford University Press,
Oxford, 2003)

\bibitem{krum}
G.\ R.\ Barsch and J.\ A.\ Krumhansl, Phys.\ Rev.\ Lett.\ {\bf 37}, 9328
(1974); G.\ R.\ Barsch, B.\ Horovitz and J.\ A.\ Krumhansl, Phys.\ Rev.\
Lett.\ {\bf 59}, 1251 (1987).

\bibitem{good}
G. S. Bales and R. J. Gooding, \prl {\bf 67}, 3412
(1991); A. C. E. Reid and R. J. Gooding, \prb {\bf 50},
3588 (1994). 

\bibitem{strn-only}G. R. Barsch {\em et al.}, \prl {\bf 59}, 1251 
(1987); K. \O.  Rasmussen {\em et al.}, \prl {\bf 87},
055704 (2001); T. Lookman {\em et al.}, \prb 67, 024114 (2003), and
references therein.

\bibitem{lqchen}
Y. Wang, L.Q. Chen, Y.H. Wen, Acta Mat., {\bf 50}, 13 (2002); L. Q. Chen,
Annu. Rev. Mater. Res. {\bf 32}, 113, (2002). 

\bibitem{rajiv}
R. Ahluwalia and G. Ananthakrishna, \prl, {\bf 86}, 4076 (2001);
S. Sreekala and G. Ananthakrishna, \prl, {\bf 90}, 135501 (2003);
S. Sreekala, R. Ahluwalia and G. Ananthakrishna, \prb {\bf 70}, 224105 (2004).

\bibitem{nishiyama}Z. Nishiyama, {\em Martensitic Transformation},
(Academic Press, NY, 1978).

\bibitem{MPP} P. C. Martin, O. Parodi, and P. S. Pershan \pra {\bf 6}, 2401 (1972).

\bibitem{ChaikinLubensky} P. Chaikin and T. C. Lubensky, {\em Principles of
Condensed Matter Physics} (Cambridge University Press, Cambridge, 1995).

\bibitem{micro}
Ph. Boullay, D. Schryvers and J.M. Ball,  Acta Mater. {\bf 51}, 1421, (2003);
D. Schryvers, P. Boullay, P.L. Potapov, R.V. Kohn and J.M. Ball, 
Int. J. of Solids and Structures, {\bf 39}, 3543, (2002)
J.M. Ball and D. Schryvers, J. de Physique IV, {\bf 112}, 159, 2003. 

\bibitem{elastop} 
V. A. Lubarda, {\em Elastoplastic theory}, (CRC Press, Boca Raton 2002). 


\bibitem{plasticity}J. Lubliner, {\em Plasticity Theory}, (Macmillan Publishing
Co., New York, 1990).

\bibitem{langer}
M. L. Falk and J. S. Langer, \pre {\bf 57}, 7192 (1998);
J. S. Langer, \pre 64, 011504 (2001);
J. S. Langer and L. Pechenik, \pre 68, 061507 (2003).

\bibitem{lemaitre}
A. Lema\^itre, \prl {\bf 89}, 195503, (2002).

\bibitem{Avad-sym}
D. M. Hatch, T. Lookman, A. Saxena, and S. R. Shenoy, Phys. Rev. B {\bf 68}, 104105 (2003)

\bibitem{toledo}
P. Toledano and V. Dmitriev, {\em Reconstructive Phase Transitions in Crystals and 
Quasicrystals}, (World Scientific, Singapore, 1996)

\bibitem{jayeetraj} J. Bhattacharya, S. Sengupta and M. Rao, preprint.

\bibitem{sq-tr1}
C. Ghatak and K.G. Ayappa, \pre {\bf 64}, 051507 (2001);
K.G. Ayappa and C. Ghatak, J. Chem. Phys. {\bf 117}, 5373 (2002);
M. Schmidt  and H. L\"owen, \pre {\bf 55}, 7228 (1997);
A. Fortini and M. Dijkstra, J. Phys.: Condens.  Matter {\bf 18}, L371 (2006);
A.H. Marcus  and S.A. Rice, \pre {\bf 55}, 637 (1996);

\bibitem{sq-tr2}C. D. Dewhurst, S. J. Levett, and D. McK. Paul, \prb
 {\bf 72}, 014542 (2005); L. Ya. Vinnikov {\it et al.},
\prb {\bf 64}, 220508 (2001); B. Rosenstein {\it et al.},
\prb {\bf 72}, 144512 (2005); S.P. Brown {\it et al.}, \prl
 {\bf 92}, 067004 (2004); R. Gilardi {\it et al.},
\prl {\bf 93}, 217001 (2004) 

\bibitem{sq-tr3}M. Rao, S. Sengupta, and R. Shankar, \prl {\bf 79},
3998, (1997); M. F. Laguna, P. S. Cornaglia, and A. A. Aligia,
\prb {\bf 69}, 104524 (2004); S. Sankararaman and R. Shankar,
\prb {\bf 67} 245102 (2003). 

\bibitem{sq-tr4}W. Wen, L. Zhang, and P. Sheng, \prl 
{\bf 85}, 5464 (2000).

\bibitem{yeth}
A. Yethiraj and A. van Blaaderen, Nature {\bf 421}, 513
(2003); A. Yethiraj {\em et al.}, \prl {\bf 92},
058301 (2004). 



\bibitem{still} T. A. Weber and F. H. Stillinger, \pre {\bf 48}, 4351 (1993)

\bibitem{dynam} The relaxation time of the Nos\'e-Hoover thermostat\cite{ums}
$Q$ determines how fast the system relaxes to the ambient temperature. 
Equilibrium properties are unaffected by this parameter, which is chosen to 
produce fast relaxation, at the same time avoiding numerical instabilities. On the other hand,
dynamical properties can be sensitive to this parameter, however,
our results are robust to changes of $Q$ within reasonable limits.  

\bibitem{ho}
J.R. Morris and K.M. Ho, \prl {\bf 74}, 940 (1995).

\bibitem{note-quench}This apparently 
unusual quench protocol needs some explanation\cite{jpcm}. The atomistic interactions 
in our model are themselves coarse-grained over much faster degrees of 
freedom, e.g., electrons and atomic magnetic (or electric) moments. In a 
real system, these fast degrees of freedom produce (in general) many-body
interactions which finally determine structure. For example, in Fe the 
face-centered to body-centered cubic structural transition is driven by 
a ferromagnetic transition\cite{hasen}. The dynamics
of these degrees of freedom, the atomic magnetic moment in the case of Fe, 
is several orders of magnitude faster than the typical timescales of atomic 
motion. Isothermal transformation of steel\cite{bhad} achieves exactly the 
conditions we have in our MD simulations, where the transformation proceeds at
a fixed temperature, which in turn fixes the 
magnitude of the magnetic interactions in the Fe atoms.
The three-body term may also couple to 
external fields (e.g., an electric field), in which case isothermal 
transformations may also be induced by changing the field. Such 
a mechanism has been reported for colloidal crystals\cite{yeth}.


\bibitem{note-OPjump} When the anisotropic parameter $\alpha = 0$, the product 
rhombic lattice is close to being triangular, 
corresponding to a strongly first order
structural transition 
with a large jump 
in the OP ($\Delta e_3 \sim \pm .3$) and a relatively large volume change\cite{jpcm,ourprl}. 
This jump in the OP at the structural transition can be made arbitrarily small by taking 
$\alpha >0$ \cite{jpcm}; our qualitative results are exactly the same, as long as 
$\Delta e_3 > 0$ at the transition.

\bibitem{edwards}S. F. Edwards and D. V. Grinev,{\em Jamming and rheology}, 
Eds. A. Liu and S. R. Nagel (Taylor and Francis, New York, 2001) 

\bibitem{liunagel}A. J. Liu and S. R. Nagel,  Nature {\bf 396}, 21 (1998).

\bibitem{landau}
L. D. Landau and E. M. Lifshitz, {\em Theory of Elasticity}, (Pergamon, 
Oxford, 1986).

\bibitem{reversibility}A. Paul, S. Sengupta and M. Rao, in preparation.

\bibitem{kaunat}
Finally, this question, and consequently reversibility, is determined by the 
symmetry of the OP strains, 
see for example, K. Bhattacharya, S. Conti, G. Zanzotto and J. Zimmer, 
Nature, {\bf 428}, 55 (2004).


\bibitem{zipp}
A. Zippelius, B. I. Halperin and D. R. Nelson, \prb {\bf 22}, 2514 (1980).

\bibitem{bhadeshia}H.K.D.H. Bhadeshia, private communication.

\bibitem{ostwald}W. Ostwald, Z. Phys. Chem., {\bf 22}, 289. (1897);
I. N. Stranski and D. Totomanov, Z. Phys. Chem., {\bf 163}, 399. (1933).

\bibitem{osy4} P. R. ten Wolde and D. Frenkel, Science 277, 1975 (1997).  

\bibitem{rttt}
M. Bouville and R. Ahluwalia, 
\prl {\bf 97}, 055701 (2006). 

\bibitem{abram}
M. Abramovitz and I. A. Stegun, {\em Handbook of Mathematical Functions},
(Dover publications, New York, 1972); W. H. Press, S. A. Teukolsky, 
W. T. Vetterling and B. P. Flannery, {\em Numerical Recipes: The Art of 
Scientific Computing} (Cambridge University Press, Cambridge, 1992). 

\end{thebibliography}
\end{document}